\documentclass[a4paper,prd,twocolumn,showpacs,superscriptaddress,floatfix,nofootinbib]{revtex4-2}

\usepackage{preamble}
\usepackage{orcidlink}
\usepackage{lipsum} \usepackage{graphicx} \usepackage{epsf} \usepackage{epsfig}
\usepackage{amssymb,amsmath,amsfonts} 
\usepackage{amssymb} \usepackage{times} \usepackage{bm}
\usepackage{pdfpages}
\makeatletter
\AtBeginDocument{\let\LS@rot\@undefined}
\makeatother

\usepackage{hyperref} \hypersetup{
colorlinks=true,        
linkcolor=blue,         
citecolor=cyan,         
}

\usepackage{graphicx}  
\usepackage{dcolumn}   
\usepackage{bm}        
\usepackage{amsfonts,amsmath,amssymb,mathrsfs}
\usepackage{color}
\usepackage{natbib,times}
\usepackage{mathtools}
\usepackage{xspace}

\newcommand{\fuka}{\texttt{FUKA}\xspace}

\newcommand{\cf}{cf.,~}
\newcommand{\ie}{i.e.,~}
\newcommand{\eg}{e.g.,~}

\begin{document}
\title{Black hole--neutron star binaries with high spins and large mass asymmetries:\\
	II. Properties of dynamical simulations}
\author{Konrad Topolski\, \orcidlink{0000-0001-9972-7143}}
\affiliation{Institut f\"ur Theoretische Physik, Goethe Universit\"at,
Max-von-Laue-Str. 1, 60438 Frankfurt am Main, Germany}

\author{Samuel D. Tootle\, \orcidlink{0000-0001-9781-0496}}
\affiliation{Department of Physics, University of Idaho, Moscow, ID 83843, USA}
\affiliation{Institut f\"ur Theoretische Physik, Goethe Universit\"at,
Max-von-Laue-Str. 1, 60438 Frankfurt am Main, Germany}

\author{Luciano Rezzolla\, \orcidlink{0000-0002-1330-7103}}
\affiliation{Institut f\"ur Theoretische Physik, Goethe Universit\"at,
Max-von-Laue-Str. 1, 60438 Frankfurt am Main, Germany}
\affiliation{Frankfurt Institute for Advanced Studies, Ruth-Moufang-Str. 1,
60438 Frankfurt, Germany}
\affiliation{School of Mathematics, Trinity College, Dublin 2, Ireland}

\date{\today}

\begin{abstract}
Black hole (BH) -- neutron star (NS) binary mergers are not only strong
sources of gravitational waves (GWs), but they are also candidates for
joint detections in the GW and electromagnetic (EM) spectra. However, the
possible emergence of an EM signal from these binaries relies crucially
on the fate of the NS which, in turn, is determined by a complex
combination of the equation of state (EOS), the BH spin, and the
mass ratio. Depending on these binary parameters, the NS can either
undergo a ``tidal disruption'' or ``plunge'' directly into the BH without
losing its compact structure beforehand. Accurate predictions as to which
of these scenarios actually takes place relies on the use of
numerical-relativity simulations, which represent essential tools to
learn about the properties of the system.
In this second paper in a series, we present a systematic exploration of
the possible space of binary parameters in terms of the mass ratio and BH
spin so as to construct a complete description of the dynamical processes
accompanying a BHNS binary merger. This second work relies not only on
the initial data presented in the companion paper I, but also on the
predictions via quasi-equilibrium (QE) sequences on the outcome of the
binary. In this way, and for the first time, we are able to relate the
predictions of QE analyses with the results of accurate
general-relativistic magnetohydrodynamic simulations.
In addition to a careful investigation of the evolution of the BH mass
and spin as a result of the merger, the total remnant rest-mass of the
resulting accretion disk and its properties, and of the corresponding
post-merger GW emission, special attention is paid to the conditions that
lead to tidal disruption. Leveraging QE calculations, we are able to
verify the reliability of stringent predictions about the occurrence or
not of a plunge and to measure the ``strength'' of the tidal disruption
when it takes place.
Finally, using a novel contraction of the Riemann tensor in a locally
orthonormal tetrad and with respect to the one comoving with the fluid
introduced in paper I, we are able to point out the onset of the
instability to tidal disruption. This new diagnostic can be employed not
only to determine the occurrence of the disruption, but also to
characterize it in terms of the binary parameters.
\end{abstract}
\maketitle

\section{Introduction}
\label{sec:introduction}

Black hole (BH) -- neutron star (NS) mergers are one of the promising
compact binary coalescence events that are targets of concurrent
contemporary gravitational-wave (GW) and electromagnetic (EM)
searches. In the O3 run of the LIGO-Virgo Collaboration, two such
candidate events have been detected~\cite{Abbott2021}, with no follow-up
EM signal. Recently, the detection of GW230523~\cite{LIGOScientific2024}
highlighted the possibility for the existence of a mixed binary system
with a BH in the range of $3-5\, M_{\odot}$. The detection of an EM
signal from a BHNS merger similar to the one from the neutron
star--neutron star (NSNS) binary system
GW170817~\cite{LIGOScientific:2017vwq}, would provide a vast amount of
information about the evolution of these systems and important
correlation between the binary parameters and the set of admissible EM
counterparts that are expected in a BHNS merger~\cite{Colombo2023}.

In the case of BHNS mergers, and with the exception of EM precursors, \ie
transients occurring prior to merger, the possible EM signal relies
crucially on the fate of the companion NS. Depending on the equation of
state (EOS) and the parameters of the binary, such as the BH
dimensionless spin $\chi_{_{\rm BH}}$, stellar compactness $C$ and the
mass ratio $q := M_{_{\rm NS}}/ M_{_{\rm BH}}\leq 1$, the NS can either
undergo a tidal disruption outside of the innermost stable circular orbit
(ISCO) of the BH or ``plunge'' directly into the BH without losing its
compact structure beforehand~\cite{Pannarale2010, Foucart2020a,
  Kyutoku2021a, Duez2024}. Accurate predictions of the outcome rely
heavily on the use of numerical-relativity simulations of these events.

To this end, this represents the second in a series of papers exploring
BHNS mergers with substantial mass asymmetry $Q := q^{-1} \geq 4$ and
high BH spins $\chi_{_{\rm BH}}=0.8$. In particular, in this work we
carry out computationally demanding general-relativistic
magnetohydrodynamic (GRMHD) simulations of BHNS mergers by combining high
quality initial data obtained with the \fuka solver
~\cite{Papenfort2021b, Tootle2024a} and highly accurate numerical methods
for the subsequent time evolution in the 3+1 spacetime
split~\cite{Most2019b}. Hence, this work (hereafter paper II) complements
the study of initial-data sequences presented in
Ref.~\cite{Topolski2024b} (hereafter paper I) and anticipates the
analysis of the ejected matter, its dynamics, but also its
nucleosynthetic and EM yields that will be presented in a subsequent
paper III~\cite{Topolski2024d}.

Numerical investigations of these systems are not new and have been
routinely conducted by a number of numerical relativity groups~\cite[see,
  \eg][]{Etienne2007b, ShibataTaniguchi2008, Foucart2011, Foucart2020b,
  Chaurasia2021, Khamesra2021, Tsao2024, Chen2024, Martineau2024,
  Matur2024}. Recent advances in our understanding of these systems are
primarily concerned with the possibility of an EM signal accompanying the
emission of gravitational waves. In this respect, the pre-merger twisting
of magnetic-field lines in the common magnetosphere might lead to the
formation and emission of plasmoids. Subsequent EM transients \eg in the
form of fast radio bursts, have been explored in force-free or resistive
magnetohydrodynamical (RMHD) frameworks by~\cite{East2021, Carrasco2021,
  Most2023a}. Pre-merger flares caused by the shattering of NS crust
represent another possibility for an EM precursor~\cite{Neill2021}.

On the other hand, the post-merger configuration involving a spinning BH
and an accretion disk can potentially launch a relativistic jet via the
Blandford-Znajek mechanism~\cite{Blandford1977}, resulting in a coherent
and directed outflow of matter in an environment dominated by magnetic
pressure. Such surroundings accelerate charged particles to significant
Lorentz factors, giving rise to a non-thermal EM emission mechanism
observable in the form of a gamma-ray burst. Since substantial magnetic
field strengths are needed to produce an impact both during the
inspiral~\cite{Giacomazzo:2009mp} and after the merger, some recent works
focus specifically on the mechanism of magnetic-field
amplification~\cite{Izquierdo2024}. Finally, the most challenging efforts
to date involve carrying out the evolution for tens to hundreds of
milliseconds \cite{Rezzolla:2010, Paschalidis2014, Ruiz2018, Most2021a}
to several seconds~\cite{Hayashi2022, Gottlieb2023a, Hayashi2023},
oftentimes at a tremendous computational cost. In return, however, this
ab-initio and self-consistent approach offers the most faithful insight
into the process of magnetic winding, angular-momentum transport in the
disk, and associated magnetic-field amplification, so as to assess the
role of magnetic fields in generating magnetically driven outflows
\citep{Kiuchi2012b, Siegel2013, Kiuchi2017, Fernandez2018, Ciolfi2020_a,
  Fujibayashi2023}, but also in producing the conditions necessary for
jet formation and launching ~\citep{Liu:2008xy, Anderson2008,
  Rezzolla:2011, Palenzuela2013a, Kiuchi2015, Murguia-Berthier2016,
  Ciolfi2020c, Nathanail2020b, Nathanail2020c, Gottlieb2022}.

Building on the results of paper I, this work has several goals and
presents a number of findings. First, it explores the dynamics of BHNS
binaries across a range of mass ratios $q$ and BH spins $\chi_{_{\rm
    BH}}$ in a physically relevant and numerically challenging
regime. Secondly, the use of high-order evolution methods allows us to
form meaningful predictions for the amount of bound and unbound material
left after the merger, the properties of the remnant BH, the
characteristics of the GW signal, and to quantify the impact of varying
the initial conditions on the nature of the event as a ``plunge'' or a
``tidal disruption''. Some of our simulations study some of the most
extreme configurations considered in the literature~\cite{Foucart2011,
  Foucart2013a, Foucart2014, Kyutoku2015}, thus allowing for a more
quantitative comparison, while also investigating novel diagnostic
quantities or employ established ones on previously neglected parts of
the parameter space. Finally, we provide for the first time a direct
correlation of predictions derived from sequences of quasi-equilibrium
(QE) initial data to quantitative and qualitative features computed in
the dynamical evolution.

The paper is structured in the following manner. In
Sec.~\ref{sec:numerical_setup}, we define the numerical setup of our
simulations, including the chosen EOS, coverage of the parameter space,
as well as the choice of evolution and gauge schemes. Moreover, we define
the necessary diagnostic quantities and discuss the level of constraint
violations throughout our simulations. Next, we describe the observed
dynamics of the merger concentrating on two different, but representative
cases of a ``plunge'' and ``tidal disruption'' in
Sec.~\ref{sec:results_general}, followed by a general discussion of the
properties of the remnant disks and BH properties in
Sec.~\ref{sec:results_remnant} for all binary configurations explored in
this study. The analysis of the extracted GW signal is presented in
Sec.~\ref{sec:results_GW}, which includes the discussion of the ringdown
signal and the frequency spectrum. Finally, we summarize the main results
in Sec.~\ref{sec:summary} and outline our future goals and possible
avenues that could be explored. In Appendix ~\ref{sec:qe_sequences}, we
provide further details regarding the comparison with quasi-equilibrium
sequences that were investigated in paper I. Throughout the paper, we
will use the geometrical units $G = c = 1$, where $G$ and $c$ are the
gravitational constant and the speed of light, respectively. Greek
(Latin) indices run from 0 (1) to 3.

\section{Numerical setup}
\label{sec:numerical_setup}

In this section, we describe the numerical setup and methods used for our
simulations. We use the infrastructure of
\texttt{EinsteinToolkit}~\cite{Loffler:2011ay} with the fixed-mesh
box-in-box refinement framework \texttt{Carpet}~\cite{Schnetter:2003rb}
to simulate eight BHNS configurations, one of which is considered with
and without the eccentricity-reducing approach discussed in
Ref.~\cite{Papenfort2021b}. All simulations are performed with three sets
of refinement-level hierarchies, centred at both objects and at the grid
origin, each consisting of seven refinement levels. An extra, 8th
refinement level is placed around both of the objects. The finest
grid-spacing around the BH and the NS is thus $147\,{\rm m}$. The total
domain extent is $(3025\, {\rm km})^3$ and reflection symmetry with
respect to the $z=0$ plane is imposed. We additionally set the
Courant--Friedrichs--Lewy (CFL) factor to $0.45$ on the inner refinement
levels, and decrease it consecutively by a factor of $2$ on the three
outermost refinement levels.

To obtain the initial configurations at the separations given in
Tab.~\ref{tab:ID_properties}, we use the initial-data solver \fuka
\cite{Papenfort2021b, Tootle2024a} based on the \texttt{Kadath}
spectral-solver library~\cite{Grandclement09}, which solves the
constraint equations on an initial hypersurface cast in the XCTS
form~\cite{Pfeiffer:2002iy, York99} under the assumption of
quasi-equilibrium, \ie the existence of global helical symmetry. \fuka
has quickly become a valuable community resource due to its ability to
reliably explore astrophysically relevant binary configurations across
many NR applications \cite{Most2021a, Papenfort:2022ywx, Kuan2023,
  Kuan2023a, Markin2023, Rosswog2023, Chen2024, Izquierdo2024,
  Corman2024, Topolski2024b}, including a recent extension to explore
scalar/tensor theories of relativity \cite{Kuan2023}. In order to
guarantee the lowest possible level of constraint violations for the
initial data, we generate our configurations using a rather high spectral
resolution of $N = \left( 17, 17, 16 \right)$ collocation points
respectively in the radial, polar and azimuthal directions, in each
domain of the numerical grid.

The orbital angular velocity $\Omega$ of the quasi-equilibrium
initial-data solution is a degree of freedom that is predominantly
constrained by enforcing the so-called ``force-balance'' at the stellar
center, \ie $\partial_i \ln h = 0$~\cite{Papenfort2021b,
  Gourgoulhon-etal-2000:2ns-initial-data, Tichy11, Kyutoku2021b}, where
$h$ is the specific enthalpy~\cite{Rezzolla_book:2013}. However, the
resulting inspiral motion is known to feature a certain degree of
eccentricity~\cite{Foucart2008, Papenfort2021b, Kyutoku2021b}, typically
of the order of $e\approx 0.02-0.03$, and increasing with higher spins
and larger mass asymmetries~\cite{Foucart2008, Kyutoku2021b}. For the
BHNS binaries studied here, this will be especially visible for the case
\texttt{Q4.chi0.8}, where the eccentricity is clearly imprinted onto the
GW signal and can impact the binary parameter
inference~\cite{Ramos-Buades2020}.

For this reason, we have performed eccentricity reduction on that
dataset, reducing the eccentricity to the acceptable level of $e \approx
10^{-3}$ and reporting below the differences between the
eccentricity-reduced and the non-reduced dataset (details on iterative
eccentricity reduction and its implementation in \fuka can be found in
\cite[\eg][]{Papenfort2021b, Tichy2019, Pfeiffer:2007yz}). On the other
hand, for the other binaries in Tab.~\ref{tab:ID_properties}, the initial
orbital eccentricity is sufficiently small that it does not influence
significantly the main features of GW signal, the final-BH properties, or
the the resulting accretion disk.

The evolution of the metric variables is handled by the \texttt{Antelope}
code~\cite{Most2019b} in the form of a 4th-order, up-wind finite
differencing scheme. We employ either the moving-punctures based
CCZ4~\cite{Alic:2011a, Alic2013} or Z4c~\cite{Bernuzzi:2009ex,
  Hilditch2012} formulation with damping coefficients $\kappa_{1}=0.02$
and $\kappa_{2}=0$, or alternatively the BSSN
formulation~\cite{Nakamura87, Shibata95, Baumgarte99} for runs with
greater mass asymmetry $Q = [6, 7]$. More specifically, for the CCZ4
formulation, we rescale the $\kappa_{1}\rightarrow \kappa_{1}/\alpha$ to
ensure stability in the presence of a BH~\cite{Alic2013}. To solve the
GRMHD equations, we use the Frankfurt-IllinoisGRMHD (\texttt{FIL})
code~\cite{Most2019b}, based on the original \texttt{IllinoisGRMHD}
code~\cite{Etienne2015} and extended with the tabulated EOS support and a
formal 4th-order scheme for the computation of numerical GRMHD
fluxes~\cite{DelZanna2007}. The quasi-local measurements and the
localization of the apparent horizon are performed by the
\texttt{EinsteinToolkit} thorns \texttt{QuasiLocalMeasures} and
\texttt{AHFinderDirect}, respectively.

Before leaving this section it is useful to comment on why some of the
simulations have not been performed with Z4-based formulations of the
Einstein equations. Such formulations, in fact, have become increasingly
common because of their constraint-damping properties, which are instead
absent in the BSSN formulation (see however \eg \cite{Yoneda02a,
  Raithel2022a, Etienne2024} for possible improvements). Our choice for
the BHNS binaries with $Q = [6, 7]$ has been motivated by the fact that
the use of the BSSN formulation leads in these cases to a better
conservation of the BH irreducible mass $M_{\rm irr}$ when compared to
the Z4-system formulations. In particular, we have observed that this
difference is especially pronounced after the merger, and that the BSSN
formulation preserves $M_{\rm irr}$ well without requiring prohibitively
high spatial resolution. At the same time, we have found that similar
drifts in the irreducible mass (which decreases) or the spin (which
increases) are present also when using the implementation of the CCZ4
formulation with the \texttt{McLachlan} thorn (which is part of the
\texttt{EinsteinToolkit}) and within the \texttt{GRChombo}
infrastructure~\cite{Clough2015}. Finally, the poorer constraint-damping
properties of the BSSN formulation -- which are particularly enhanced by
the NS movement across the grid and advection of constraint-violating
modes of the order of $10^{-6}-10^{-5}$ near the refinement-level
boundaries, have been compensated by the use of high-resolution initial
data and the adoption of high-order finite-differencing schemes for both
metric and hydrodynamical variables. As a result, the constraint
violations for the BSSN runs after the merger are of the order of
$10^{-7}$ across the entire numerical grid and thus reasonably small
overall. It is possible and likely that similarly small violations of the
irreducible mass conservation can be obtained also with the Z4-system formulations by
increasing the resolution, as observed in the case of NSNS
binaries~\cite{Alic2013, Kastaun2013}, but this needs to be verified in a
future work.

\subsection{Space of parameters}
\label{subsec:param_space}

We now move to discuss the binary configurations that are evolved in this
work. First, for the EOS needed to close the system of GRMHD equations we
have employed the DD2 EOS~\cite{Typel:2009sy, Hempel:2009mc}, whose
self-consistent temperature and composition dependence allows us to
analyze the temperature and composition of nuclear matter during the
merger and post-merger phases. Furthermore, for a meaningful comparison
among configurations with different BH spins and mass ratios, we fix the
gravitational mass of the NS to be $M_{_{\rm NS}}=1.4\, M_{\odot}$ and
the corresponding baryonic rest mass $M_{\rm b}=1.53\, M_{\odot}$. The
proper radius of $R_{_{\rm NS}}=12.93\,{\rm km}$ is obtained by solving
for the NS in isolation and fixing the Arnowitt-Deser-Misner (ADM) mass
of the (nonrotating) NS to $M_{_{\rm NS}}$. While this radius is somewhat
larger than the one expected from agnostic considerations about the sound
speed in NSs and which predict that $R_{_{\rm NS}} =
12.42^{+0.52}_{-0.99} \, {\rm km}$~\cite{Altiparmak:2022}) with a 95\%
confidence level, this is actually advantageous as it favours tidal
disruption, enhancing an important aspect of our
analysis. Table~\ref{tab:ID_properties} summarizes the initial binary
parameters and reports the important properties such as: the inverse mass
ratio $Q$, the dimensionless spin of the BH $\chi_{_{\rm BH}}$, the
initial Christodoulou mass $M_{\rm Ch}$, the total gravitational mass of
the compact objects as measured in isolation $M_{\rm tot}$, the initial
separation $d_{0}$, and the estimated eccentricity $e$. For completeness,
we recall that the Christodoulou mass is defined here as
\begin{align}
  M_{\rm Ch}^2 := M_{\rm irr}^2 + \mathcal{S}^2 / 4M_{\rm irr}^2 \,;
  \nonumber
\end{align}
where $M_{\rm irr}$ is the irreducible mass of the BH and $\mathcal{S} =
M_{\rm Ch}^{2}\chi_{_{\rm BH}}$ is the BH spin angular momentum, which we
measure on the apparent horizon.

\begin{table}[t]
  \begin{ruledtabular}
    \begin{tabular}{l|cccrccc}
      binary & $Q$ & $\chi_{_{\rm BH}}$ & $M_{\rm Ch}$ &
      $M_{\rm tot}$ & $d_{0}$ & $e$ & scenario\\
	    &&& $[M_{\odot}]$ & $[M_{\odot}]$ & $[M_{\odot}]$ &
	    $[10^{-2}]$ & \\
      \hline
      \texttt{Q4.chi0.0}    & $4$ & $0.0$ & $5.6$ & $7.0$  & $56$ &  --   & PL\\
      \texttt{Q4.chi0.4}    & $4$ & $0.4$ & $5.6$ & $7.0$  & $56$ & $3.2$ & TD \\
      \texttt{Q4.chi0.6}    & $4$ & $0.6$ & $5.6$ & $7.0$  & $56$ & $2.8$ & TD\\
      \texttt{Q4.chi0.8}    & $4$ & $0.8$ & $5.6$ & $7.0$  & $56$ & $2.5$ & TD\\
      \texttt{Q5.chi0.8}    & $5$ & $0.8$ & $7.0$ & $8.4$  & $56$ & $3.7$ & TD\\
      \texttt{Q6.chi0.8}    & $6$ & $0.8$ & $8.4$ & $9.8$  & $60$ & $4.6$ & TD\\
      \texttt{Q7.chi0.8}    & $7$ & $0.8$ & $9.8$ & $11.2$ & $64$ & $5.5$ & TD\\
      \hline
      \texttt{Q4.chi0.8.er} & $4$ & $0.8$ & $5.6$ & $7.0$  & $56$ & $0.8$ & TD
    \end{tabular}
  \end{ruledtabular}
  \caption{Properties of the initial data for the simulations in this
    work. Reported are: the inverse mass ratio $Q:= q^{-1} =M_{_{\rm Ch}}
    / M_{\rm NS}$, the initial BH spin $\chi_{_{\rm BH}}$, the initial
    separation $d_{0}$, the BH Christodoulou mass $M_{\rm Ch}$, and the
    estimated eccentricity $e$ based on the dynamical evolution. The
    outcome of the merger as a tidal disruption (TD) or plunge (PL) is
    reported in the last column.}
\label{tab:ID_properties}
\end{table}

Because we are mostly interested in exploring the dynamics of matter
and the GW emission in the two most interesting scenarios produced by
BHNS mergers, \ie a ``tidal disruption'' and a ``plunge'', our study
concentrates on two main sets of binary configurations, which we discuss
separately below, and that assess the impact that the mass ratio and BH
spin have on the possibility of tidal disruption. The first set of the
eight BHNS configurations evolved here includes four binaries with a fixed
mass ratio $q = 1/4$ and where we vary the BH spin in the range
$\chi_{_{\rm BH}} = [0,0.4, 0.6, 0.8]$. As a side remark, we note that
the largest BH spin of $\chi_{_{\rm BH}}=0.8$ is close to the limit
achievable under the assumption of conformal flatness in the initial
data, \ie $\chi_{_{\rm BH}} \approx 0.85$~\cite{Papenfort2021b,
  Lovelace2008c}. Furthermore, since $q \leq 1$ under realistic
astrophysical conditions and to simplify our notation, we will hereafter
indicate the mass ratio as $Q \geq 1$.

\begin{figure}
  \centering
  \includegraphics[width=1.00\columnwidth]{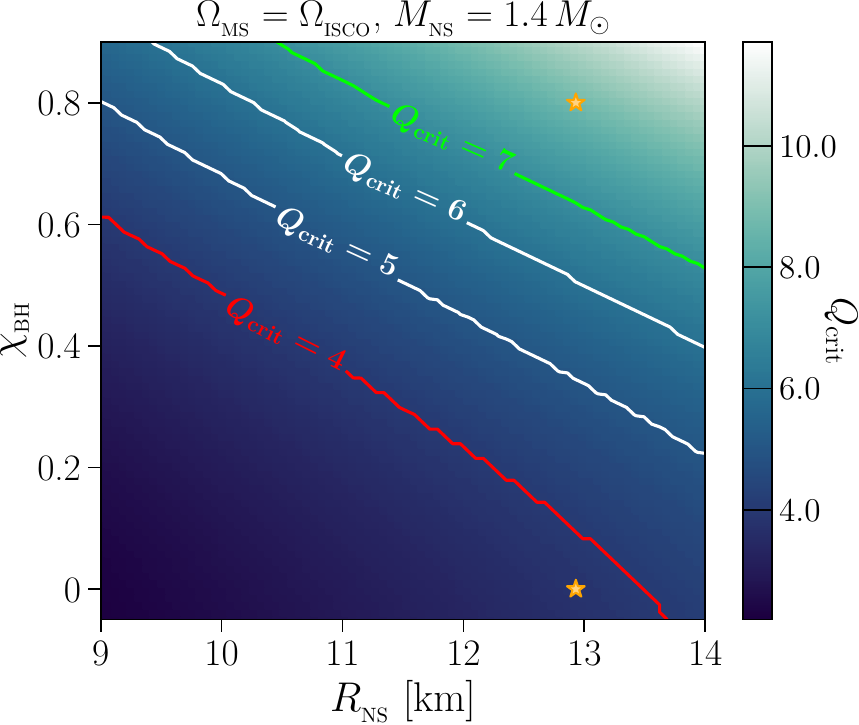}
  \caption{Critical mass ratios $Q_{\rm crit}$ derived from the
    disruption condition $\Omega_{_{\rm ISCO}}=\Omega_{_{\rm MS}}$ for a
    fixed NS mass of $M_{_{\rm NS}}=1.4 \, M_{\odot}$. The dependence in
    the $(R_{_{\rm NS}}, Q, \chi_{_{\rm BH}})$ space is based on QE
    sequences with DD2 EOS (see Fig.~13 in paper I). For a fixed $Q_{\rm
      crit}$, scenarios with lower NS radii and BH spins lead to a plunge
    event; higher NS radii and BH spins instead favor tidal
    disruption. Golden stars depict two representative binaries:
    \texttt{Q4.chi0.0} and \texttt{Q7.chi0.8}. The former is below the
    corresponding $Q_{\rm crit}$ line (red contour) and leads to a
    plunge; the latter is above the corresponding $Q_{\rm crit}$ line
    (green contour) and leads to a tidal disruption.}
  \label{fig:Om_Dis_Cond_DD2}
\end{figure}

For the second set of simulations, on the other hand, we fix the BH
dimensionless spin to $\chi_{_{\rm BH}} = 0.8$, but increase the mass
ratio to more realistic values of $Q=5, 6, 7$. This set includes binaries
in the region of parameter space where population synthesis studies
suggest a peak in the distribution of BHNS mass
ratios~\cite{Broekgaarden2021}, and which result in more realistic
remnant black hole masses in the range of $M^{\rm
  rem}_{_{\rm_{BH}}}\approx 7-10 \, M_{\odot}$~\cite{Zappa2019}.

Additionally, we endow the NS star with a poloidal magnetic field via the
vector potential
\begin{align}
  A^{i} &= A_{0}\max\{p-p_{\rm cut},0 \}^{n} \\
  &\phantom{=}\times [(x-x_{\rm
      NS})\delta^{i}_{y} - (y-y_{\rm NS})\delta^{i}_{x}]  \,, \nonumber
\end{align}
with parameters $n=2$, $p_{\rm cut}=10^{-6}$, $A_{0}=300$ and where
$x_{\rm NS},y_{\rm NS}$ are coordinates of the NS centre. The resulting
magnetic-field in the NS and in the post-merger accretion disk is
moderate, albeit still unrealistically high for an old neutron star in a
binary, with an average strength $\vert B \vert \approx 10^{13}\, {\rm
  G}$ before and $\vert B \vert \approx 10^{12}\, {\rm G}$ after the
merger. Without a suitable magnetic-field amplification mechanism
operating, it has negligible impact on processes occurring over a few
dynamical timescales, as those discussed here.

The initial separation in the binary components varies as a function of
$Q$ to minimize the computational cost of the simulation while ensuring
that a sufficient number of orbits are obtained prior to merger. In this
way, the binary has sufficient time to reach equilibrium within the
evolution gauge and a faithful calculation of the merger and post-merger
dynamics, and of the resulting GW signal, is possible. More specifically,
the initial coordinate distances between the two objects are chosen to be
$d_0/{\rm km} \approx [85, 85, 89, 95]$ (or, equivalently, $d_0 / M_{\rm
  tot} \approx [8.0, 6.7, 6.2, 5.7 ]$) for the cases with $Q=[4, 5, 6,
  7]$, respectively. In addition, for configurations including BHs with
prograde spins, for which the angular-momentum reservoir to radiate is
larger, the same number of orbits can be obtained with a smaller
separation (see Tab.~\ref{tab:ID_properties}).

Our choice of mass ratio has also been informed by the detailed analysis
carried out in paper I, specifically regarding the separatrix between
plunge and tidal-disruption events. In particular, we report in
Fig.~\ref{fig:Om_Dis_Cond_DD2} a representation of the separatrix between
a ``plunge'' and ``tidal disruption'' regimes for the DD2 EOS considered
here (see the analogous but more general Fig.~13 in paper I, which refers
instead to three generic EOSs). We recall that the orbital angular
velocities $\Omega_{_{\rm MS}}$ and $\Omega_{_{\rm ISCO}}$ represent the
predictions of QE sequences, and correspond to the onset of mass shedding
and the crossing of the innermost stable circular orbit,
respectively. Setting these two frequencies to be equal allows one to
find a critical mass ratio $Q_{\rm crit} = q^{-1}_{\rm crit}$, which is
reported with a colormap in Fig.~\ref{fig:Om_Dis_Cond_DD2}. Stated
differently, for a fixed NS radius and BH spin, mass asymmetries larger
than $Q_{\rm crit}$ lead to a plunge, and smaller ones favour a tidal
disruption. The white lines correspond to isocontours of $Q_{\rm crit}$,
while the red (green) line refers to $Q_{\rm crit}=4$ ($Q_{\rm crit}=7$),
and is the relevant contour for our sequences with varying BH spin (mass
ratio) and with $Q=4$ ($Q=7$). We also recall that, as highlighted in
paper I, at a fixed mass ratio $Q_{\rm crit}$, smaller NS radii and lower
BH spins (down and to the left of the contour) lead to a plunge, while
larger NS radii with higher BH spins (up and to the right of the contour)
favour a tidal disruption. To help interpret the content of
Fig.~\ref{fig:Om_Dis_Cond_DD2}, we mark with golden stars the BHNS
binaries \texttt{Q4.chi0.0} and \texttt{Q7.chi0.8}. For the former, we
expect (and observe) a plunge, while for the latter we expect (and
observe) a tidal disruption. In light of these considerations,
restricting to $Q=7$ as the most asymmetric BHNS binary ensures that we
probe the critical region of tidal disruption for the compactness of
$\mathcal{C}=0.16$ for our DD2 EOS (\ie $\Omega_{_{\rm MS}} \lesssim
\Omega_{_{\rm ISCO}}$). Furthermore, as we will demonstrate later, all
the binaries considered here with rapidly spinning black holes (\ie
$\chi_{_{\rm BH}} = 0.8$) do indeed lead to a tidal disruption. The fact
that the results of QE sequences yield predictions that are consistent
with the dynamical evolutions is all the more impressive when considering
the information that they do not take into account, namely, the radiative
losses of energy and angular momentum via GWs.

It may be useful to remark that the BHNS binaries with $Q=6, 7$ and
$\chi_{_{\rm BH}}=0.8$ represent some of the most challenging systems of
this type considered so far. Indeed, BHNS systems with mass asymmetries
larger than $Q=5$ are not simulated often, mostly because the combination
of a large BH mass and small spins is likely to lead to plunge events
that are astrophysically less interesting and leave no residual matter
outside of the remnant BH (see Fig.~\ref{fig:Om_Dis_Cond_DD2} and also
the discussion in Ref.~\cite{Foucart2013b}). Furthermore, at such large
mass asymmetries and with $\chi_{_{\rm BH}} \lesssim 0.75$, the GW signal
is very similar to the corresponding one produced by a binary black-hole
(BBH) system with the same mass and mass ratio. Overall, the eight BHNS
systems selected and evolved here in full GRMHD and with a realistic
temperature-dependent EOS represent a very comprehensive sample and hence
are useful to build the ``broad-brush'' picture of the dynamics of
matter and of the GW emission that is to be expected from these systems.

\subsection{Gauge conditions}
\label{subsec:gauge_conditions}

Due to large mass asymmetries and BH spins explored with our binaries,
the surface of the NS might become severely distorted not only because of
physical tidal effects, but also as a result of a non-optimal choice of
the spatial gauge for the shift vector (see, \eg Refs.~\cite{Lousto2010,
  Mueller:2010bu}). Under these conditions, and despite the use of grid
refinement techniques, the effective resolution in the NS can decrease
and with it the quality of the solution. For these reasons, and to
mitigate the coordinate-distortion effects, we have found it helpful to
introduce a spatially varying damping term in the Gamma-driver condition
based on the symmetry-seeking shift conditions presented in
Ref.~\cite{Alic:2010}. More concretely, we use the shift-evolution
equations cast as a first-order system in the covariant form, \ie with
the inclusion of advection terms along the shift vector
\begin{align}
  \partial_{t}\beta^{i} &= k B^{i}  + \beta^{j} \partial_{j}\beta^{i} \,, \\
  \partial_{t} B^{i}    &= \big{[} \partial_{t}\tilde{\Gamma}^{i} -
    \beta^{j}\partial_{j}\Gamma^{i}\big{]}
  + \beta^{j}\partial_{j}B^{i} - \eta B^{i} \,,
  \label{eq:gamma_driver}
\end{align}
with $\beta^{i}$ the shift vector, $B^{i}=\partial_{t}\beta^{i}$,
$k=0.75$ and $\tilde{\Gamma}^{i}$ the conformal connection vector, which
has a different definition depending on whether the BSSN~\cite{Alic:2010}
or the Z4-based evolution scheme~\cite{Alic2013} is used\footnote{In CCZ4
or Z4c scheme, the conformal connection vector is evolved with the
addition of the $\tilde{\gamma}^{ij}Z_{j}$ term, where $Z_j$ is the
spatial projection of the $\mathcal{Z}^{\mu}$ vector.}. The damping term
$\eta$ is then governed by the evolution equation
\begin{align}
  \partial_{t}\eta =  \frac{1}{M_{\rm tot}}(-\eta + S(r)) +
  \beta^{i}\partial_{i}\eta \,,
  \label{eq:dynamical_eta_evol}
\end{align}
and although Eq.~\eqref{eq:dynamical_eta_evol} includes a user-specified
factor of $1/M_{\rm tot}$ at runtime, in practice the gauge evolution
does not strongly depend on the total mass. Instead, this factor sets the
timescale for the relaxation of $\eta$ to the driving term which can be
tuned as necessary (in our experience, simply using $1/M_{\rm tot}$
proved to be robust). The general expression for the driving term $S(r)$
in turn reads
\begin{align}
  S(r) = A \frac{\sqrt{\tilde{\gamma}^{ij}\partial_{i}W
      \partial_{j}W }}{(1-W^{a})^{b}}
  \left(\frac{R^{2}}{r_{1}^{2} + r_{2}^{2} + R^{2}} \right) e^{-{r^{2}}/{R^{2}}} \,,
  \label{eq:dynamical_eta_Sofr}
\end{align}
where a conformal decomposition of the spatial metric is used such that
$\gamma_{ij}=\psi^{4} \tilde{\gamma}_{ij}$, where $\tilde{\gamma}_{ij}$
is the conformal metric, $\psi$ is the conformal factor and
$W:=\psi^{-2}$. Furthermore, $r_{1}$ and $r_{2}$ are the coordinate
positions of the two objects and $R$ is a damping radius, which we set to
the initial coordinate distance between the two compact objects. In the
driving term above, we set $a=2=b=2$, which has been shown to produce
optimal gauge speeds and numerical stability~\cite{Lousto2010}. The term
$e^{-r^2 / R^2}$, while not part of the original formulation, has been
added to ensure sufficient radial damping of the $\eta$ parameter and in
this way avoid the violation of the CFL-like condition that would lead to
numerical instabilities~\cite{Schnetter:2010cz}.

\subsection{Diagnostics and observables}
\label{subsec:sim_diag_obs}

The gravitational signal emitted from an inspiraling binary can be
extracted with the help of the Weyl (pseudo)scalar
$\psi_{4}$~\cite{Alcubierre:2006, Bishop2016}, measured on a sphere at a
specified distance which we choose to be $r_{\rm det}=600\, M_{\odot}$
for all the simulations presented here. For our analysis, however, the
two polarizations of the dimensionless GW strain $h_{+,\times}$ are the
preferred variables, related to $\psi_{4}$ by second time derivatives
(\ie $\psi_4 = \partial^2_t (h_+ - ih_{\times})$). While a number of
works have argued for $\psi_{4}$ as an analysis variable with
advantageous properties for physical insight~\cite{Bustillo2022,
  Topolski2024}, and shown equivalent performance in Bayesian inference
studies~\cite{Bustillo2022}, we do not use it here as a quantity for GW
analysis in order to facilitate comparison with existing literature.
Hence, we utilize the standard decomposition in a vector basis such that
\begin{equation}
  \label{eq:psi4}
  h_{+}+i h_\times:=
  \sum_{\ell=2}^{\infty}\sum_{m=-\ell}^{m=\ell}h^{\ell,m}
      {_{-2}Y}_{\ell,m}\,,
\end{equation}
where $_sY_{\ell,m}(\theta,\phi)$ represent spin-weighted spherical
harmonics of weight $s=-2$~\cite{Bishop2016}. In the following, we will
exclusively concentrate on the dominant $\ell=m=2$ mode and consistently
denote its polarizations with the $h_{+/\times}$ symbol. To investigate
the GW signal in the frequency domain, we follow~\cite{Takami:2014zpa}
and compute the power spectral density as
\begin{align}
  \label{eq:PSD}
  \tilde{h}(f):=\frac{1}{\sqrt{2}}
	\Bigg{(}
  &\left|\!\int\! d t\,{\rm e}^{-2\pi ift}
	h_{+}(t)\right|^2 + \nonumber\\
  &   \left|\!\int\! d t\,{\rm e}^{-2\pi ift}
	h_{\times}(t)\right|^2
	\Bigg{)}^{1/2}\,,
\end{align}
where the integration runs through the whole duration of the signal. We
will also use the instantaneous GW frequency $f_{_{\rm GW}}$ defined as
\begin{equation}
	\label{eq:fGW}
    f_{_{\rm GW}}:=\frac{1}{2 \pi} \frac{d \phi}{dt}\,,\qquad \phi:={\rm
	arctan} \left( \frac{h_{\times}}{h_{+}} \right)\,.
\end{equation}
The merger time $t_{\rm mer}$ is henceforth defined as the time when
strain amplitude $|h| := \sqrt{h_{+}^{2}+h_{\times}^{2}}$ reaches its
first maximum value. Finally, the instantaneous frequency at merger is
defined as $f_{\rm mer}=f_{\rm GW}(t=t_{\rm mer})$.

We next define the quantities crucial for evaluating the merger dynamics
and its outcome. More specifically, to measure the \textit{total remnant
  rest-mass} after the coalescence, we compute the integral outside of
the apparent horizon
\begin{align}
  M_{\rm b, rem} := \int_{r>r_{\rm AH}} \rho W \sqrt{\gamma} d^{3}x\,,
  \label{eq:rem_mass_int}
\end{align}
where $\rho$ is the rest-matter density, $W$ the Lorentz factor and
$\gamma$ the determinant of the spatial metric $\gamma_{ij}$. In general,
this quantity will include both the bound and unbound matter, \eg the
dynamical ejecta. It is useful to remark that while it is often ignored,
the bound ejected matter can also be quite substantial and comparable
with the unbound one and hence play an important role in the generation
of an EM signal~\cite{Musolino2024}.

Since we are specifically interested here in the amount of matter left
around the BH, we additionally define a volume integral over
$\mathcal{V}_{100}$, that is, restricted to a sphere of radius of
$R_{100}:= 100\,M_{\odot}\approx 147\, {\rm km}$ and compute the
rest-mass of the accretion disk as
\begin{align}
  M_{\rm disk} := \int_{\mathcal{V}_{100}} \rho W \sqrt{\gamma} d^{3}x \,.
  \label{eq:disc_mass_int}
\end{align}
As we will discuss later on, $\mathcal{V}_{100}$ at the final time of the
simulation encompasses well the disk structure and does not include the
contribution from the dynamical ejecta. An alternative approach for
estimating the disk mass involves computing the bound portion of the
matter as
\begin{equation}
M_{\rm bound} := M_{\rm b, rem} - M_{\rm ej}\,,
  \label{eq:disc_mass_bound}
\end{equation}
where $M_{\rm ej}$ is defined similarly to Eq.~\eqref{eq:rem_mass_int},
but the integrand includes only the rest-mass density of unbound matter,
\ie with $u_{t}\leq -1$\footnote{We use the \textit{geodesic} criterion
instead of the Bernoulli one $h u_{t}\leq -1$, with $h$ the fluid
specific enthalpy; see Ref.~\cite{Bovard2016} for a detailed
discussion.}. In essence, $M_{\rm bound}$ generically includes both the
proto-accretion disk and the ejected, but bound material that will 
fallback on a timescale of $\mathcal{O}(10\,{\rm ms})$. Hereafter, we
will use $M_{\rm disk}$ as the main monitored quantity, but will also
compare it with $M_{\rm bound}$.

\begin{figure*}[t!]
  \includegraphics[width=0.495\textwidth]{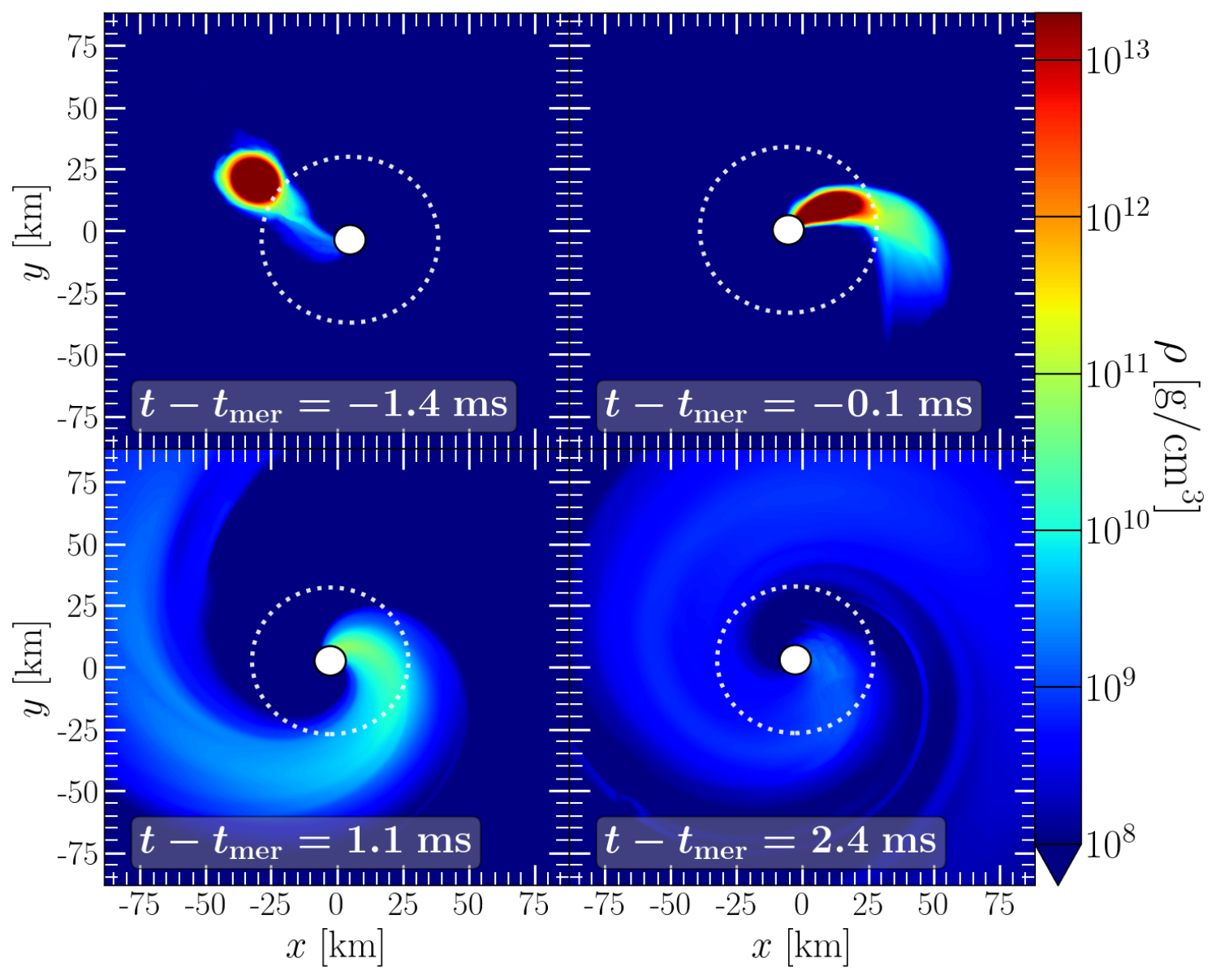}
  \includegraphics[width=0.495\textwidth]{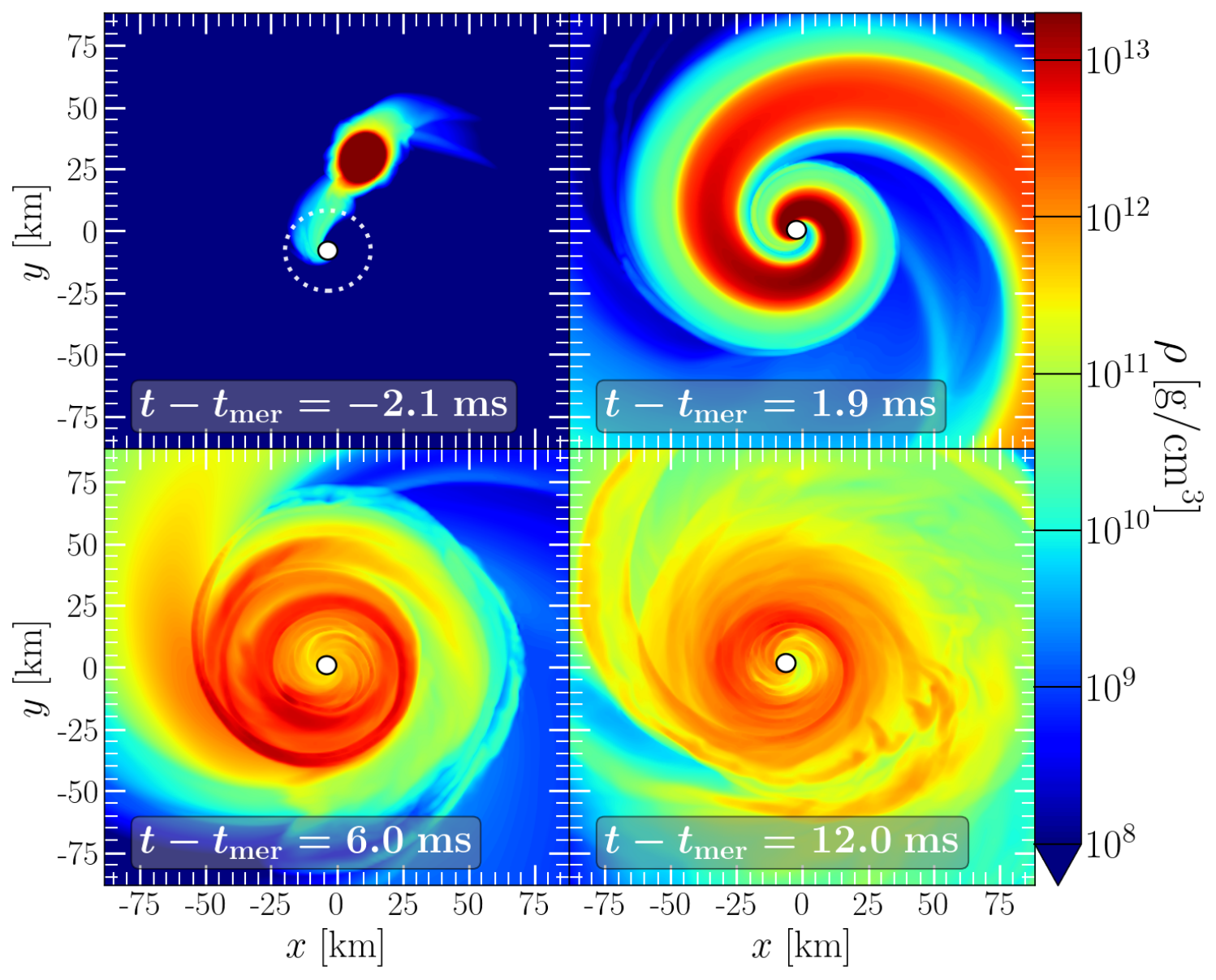}
  \caption{Representative stages of the evolution of a BHNS binary in the
    two possible scenarios of plunge or tidal disruption. \textit{Left}:
    An irrotational binary leading to a plunging NS where the remnant BH
    has mass $M^{\rm rem}_{_{\rm Ch}}=6.853\,M_{\odot}$ and spin
    $\chi^{\rm rem}_{_{\rm BH}}=0.474$. Shortly after the last frame,
    there is negligible baryonic mass outside the event horizon (filled
    white circle) left. \textit{Right}: A binary with a rapidly spinning
    BH with $\chi_{_{\rm BH}}=0.8$ resulting in a tidal disruption. The
    final BH has mass $M^{\rm rem}_{_{\rm BH}}=6.582\,M_{\odot}$, with
    spin $\chi^{\rm rem}_{_{\rm BH}}=0.868$. Outside of the apparent
    horizon (filled white circle), an accretion disk of $0.189\,
    M_{\odot}$ has formed. In all frames, dotted white circles correspond
    to $R_{\rm ISCO}$ for the Kerr metric in Boyer-Lindquist coordinates
    for the corresponding mass and spin for each BH at a relevant stage; the NS has a mass
    $M_{_{\rm NS}}=1.4\,M_{\odot}$ and $Q=4$.}
	\label{fig:dynamics_comparison}
\end{figure*}

\section{Results: general merger dynamics}
\label{sec:results_general}

In what follows we describe the most essential features of the two
possible outcomes of the BHNS merger, namely, the ``plunge'' scenario and
the ``tidal-disruption'' scenario. Furthermore we will also fix the
nomenclature necessary to describe our results for all the examined
configurations discussed here and in the remaining sections. More
specifically, we will compare and contrast the BHNS configurations
\texttt{Q4.chi0.0} and \texttt{Q4.chi0.8}, which clearly have the same
ratio $Q=4$, but where the initial BH spin is either $\chi_{_{\rm BH}}=0$
-- leading to a plunge scenario -- or $\chi_{_{\rm BH}}=0.8$ -- leading
to a disruption scenario (see also Fig.~\ref{fig:Om_Dis_Cond_DD2}, where
these two binaries are reported with golden stars).

\subsection{Plunge vs tidal-disruption}

While both of the representative binaries start from the same initial
separation $d\approx87\, \rm{km}$, thus resulting in similar orbital
angular velocities $M_{\odot} \Omega \approx 0.0053$, the binary with a
rapidly spinning BH has almost twice as much orbital angular momentum as
the irrotational case ($J_{\rm ADM}=51.492\,M_{\odot}^{2}$ and $J_{\rm
  ADM}=28.247\,M_{\odot}^{2}$). Therefore, significantly more orbital
angular momentum needs to be radiated away through GW emission during the
inspiral phase before merger, which increases the number of orbits from
$N_{\rm orb}=4$ for \texttt{Q4.chi0.0} to $N_{\rm orb}=7$ for
\texttt{Q4.chi0.8}.

\subsubsection{Plunge scenario}

We start with the irrotational BHNS binary \texttt{Q4.chi0.0}, whose
representative evolution stages are presented in the left panels of
Fig.~\ref{fig:dynamics_comparison}. Note that although this represents a
plunge scenario in which the NS is essentially accreted intact, the NS
experiences mass loss from its equatorial point closer to the BH and,
later on, from the low-density part trailing behind, which is often
referred to as the ``tidal tail''. As a result, the matter from the
largely intact NS enters the event horizon at a narrow angle (upper right
and lower left panels), leaving negligible traces of baryonic matter
outside (lower right panel). As a result of the merger, the BH mass
increases to $6.853\,M_{\odot}$ and the inflow of angular momentum spins
up the previously non-spinning BH up to $\chi_{_{\rm BH}}=0.474$. The
missing $0.146\,M_{\odot}$ lost from the initial ADM mass of the system
has been radiated in the form of GWs [see
  Fig.~\ref{fig:GW_q4567_chiBHs}], which will be analyzed in detail in
Sec.~\ref{sec:results_GW} and whose inspiral part does not differ
substantially from a GW signal produced by an equivalent BBH
configuration. Indeed, the major difference between such a BHNS and an
equivalent BBH binary during the inspiral is to be found in the phase
difference introduced by the tidal deformation of the NS, which is
measured through the tidal polarizability $\Lambda$ or the tidal coupling
constant $\kappa_{2}^{^{T}}$. Likewise, the post-merger signal is similar
but different from the corresponding signal in a BBH merger and this is
mostly because the amount of matter accreted after the NS has been
absorbed is rather small and does not impact significantly the newly
formed BH.

In summary, in the case of the \texttt{Q4.chi0.0} binary, insufficient
remnant matter is left outside the event horizon to form an accretion
disk nor is there a significant amount of material ejected during the
late inspiral. Both of these conditions reduce significantly the
possibility of a detectable EM counterpart.

\subsubsection{Tidal-disruption scenario}

The dynamics of the \texttt{Q4.chi0.8} binary, which leads to the tidal
disruption shown in Fig.~\ref{fig:dynamics_comparison} (right panels), is
much more interesting. In this case, the substantial BH spin causes
the ISCO orbit (\ie the circular orbit at which $\Omega=\Omega_{_{\rm
    ISCO}}$) to move inward. As a consequence, the stellar deformation
and mass-shedding due to large tidal forces and orbital velocities occurs
well outside of the ISCO. The formation of a cusp on the NS surface
(upper left panel) is followed by a significant tidal tail. While this
mass transfer is unstable, the motion of individual fluid elements can
often be modelled successfully by the approximation of massive geodesics
in a curved, Kerr-like spacetime~\cite{Pannarale2010, Hayashi2021}.

Leaving a more detailed discussion to paper III~\cite{Topolski2024d}, we
here note that during the tidal disruption, kinetic energy and angular
momentum are transferred from the orbital motion over to the disrupted NS
matter. In general, fluid elements in the tidally disrupted material
closer to the BH will either be captured (if at the time of disruption
they are located below the ISCO) or otherwise be part of the newly formed
accretion disk. Furthermore, fluid elements at a larger distance from the
BH will gain energy and angular momentum through transport along the
tidal tail and may become gravitationally unbound, leading to what is
normally referred to as the ``dynamical ejecta''. The resulting tidal
tail (shown in the upper right panel) winds around the BH due to its
differential angular velocity and the number of distinguishable
``spirals'' depends on the location of the onset of mass-shedding and the
instantaneous orbital angular velocity at that time, which increases for
less massive and rapidly spinning BHs. The collision of the leading edge
of the tidal tail with its trailing edge forms a massive proto-accretion
disk with a mass of $0.189\,M_{\odot}$ (two last panels) around the
remnant BH, with a mass of $M^{\rm rem}_{_{\rm Ch}}= 6.582\,M_{\odot}$
and a final spin $\chi^{\rm rem}_{_{\rm BH}}=0.868$. Finally, due to
shock heating during the collision of the front and back of the tidal
tail, the temperature of matter increases to $\approx 10\,\rm MeV$.

In paper I, we have presented a detailed analysis of a novel way of
characterising in a quantitative manner the tidal deformation experienced
by the NS before the disruption. In particular, we have introduced the
so-called ``tidal-force'' operator $\mathcal{C}_{\hat{i}\hat{j}}$,
appearing in the geodesic deviation equation for a judiciously chosen
reference frame~\cite{Marck83, Wiggins00} [see Eqs. (23) and (24) of
  paper I]\footnote{The evolution of the deviation vector components
$X^{\hat{i}}$ in a geodetic, parallel transported tetrad
$\boldsymbol{e}_{\hat{a}}$ is governed by $d^{2}X_{\hat{i}}/dt^{2} =
C_{\hat{i}\hat{j}} X^{\hat{i}}$, with $C_{\hat{i}\hat{j}} =
\boldsymbol{R} (\boldsymbol{e}_{\hat{i}}, \boldsymbol{e}_{\hat{0}},
\boldsymbol{e}_{\hat{0}}, \boldsymbol{e}_{\hat{j}})$ and $\boldsymbol{R}$
representing the Riemann tensor.}. Here, hatted indices correspond to the
indices of an orthonormal tetrad, \ie a set of locally defined unit
four-vectors. The square of this matrix $\Upsilon :=
\mathcal{C}_{\hat{i}\hat{j}} \mathcal{C}^{\hat{i}\hat{j}}$ is independent
of the spatial vectors in the tetrad and only depends on the observer
four-velocity, which we choose to be the one comoving with the fluid (see
paper I for a discussion on the choice of the tetrad). Stated
differently, the scalar function $\Upsilon$ provides a simple and yet
effective way of measuring not only the degree of deformation of the NS,
but also provides a covariant measure of the point-wise magnitude of
tidal forces.

\begin{figure}
  \includegraphics[width=0.99\columnwidth]{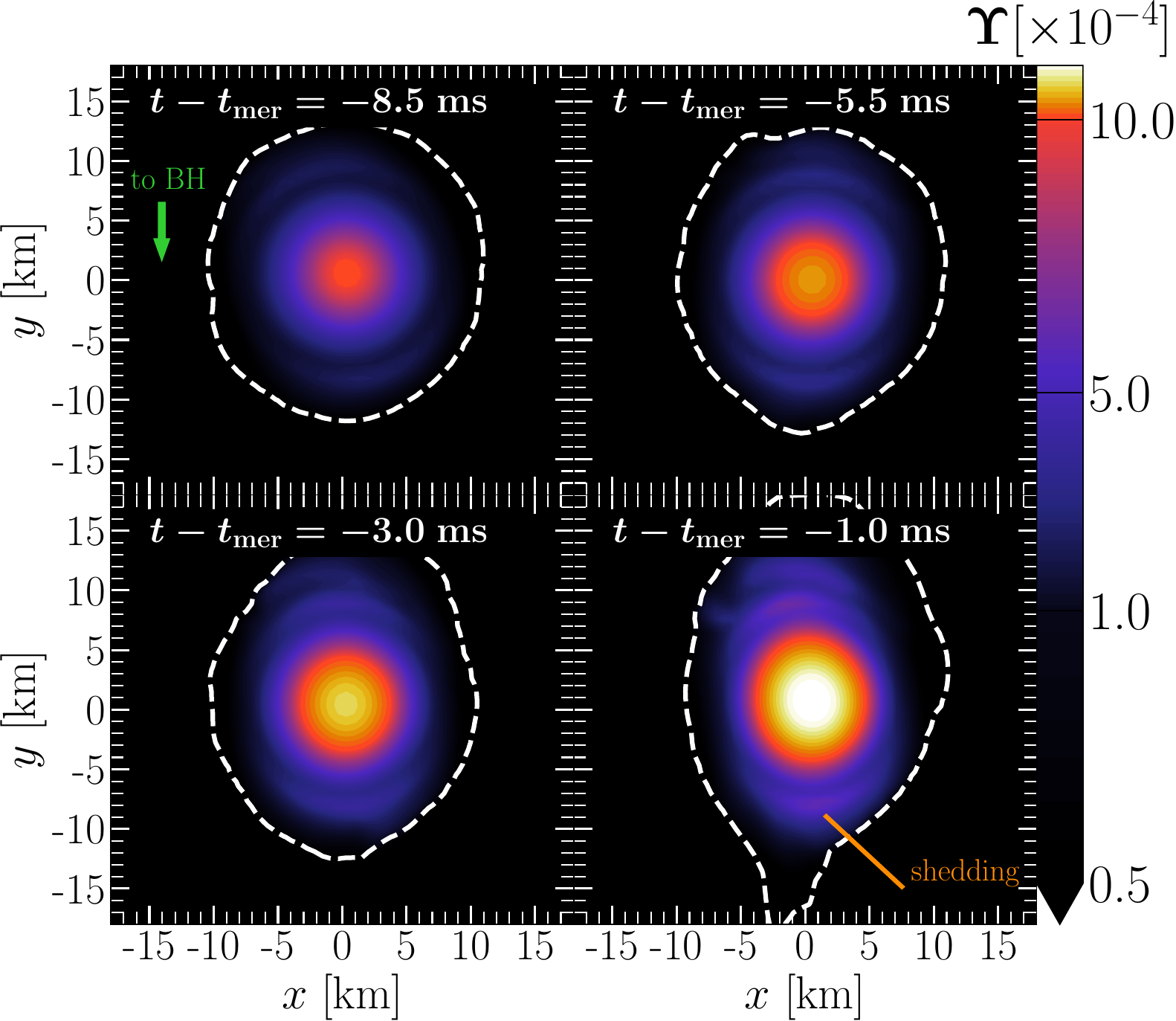}
  \caption{Representative stages of the evolution of the tidal-force
    indicator $\Upsilon=C_{\hat{i}\hat{j}}C^{\hat{i}\hat{j}}$ on the
    equatorial plane for the \texttt{Q7.chi0.8} binary. The data is
    rotated for each panel so that the green arrow always points in the
    direction of the BH. The orange line indicates the region of the star
    that will undergo mass-shedding and the dashed lines represent the
    iso-density contour corresponding to $\rho = 6.2 \times 10^{9} \,
    {\rm g}/{\rm{cm}^{3}}$. Note how the deformation increases as the
    inspiral proceeds and how it develops an anisotropy in the direction
    to the BH.}
    \label{fig:evol_q7_tidalforceindicator}
\end{figure}

This is illustrated in Fig.~\ref{fig:evol_q7_tidalforceindicator}, which
reports the evolution of this diagnostic scalar at representative times
prior to merger for the \texttt{Q7.chi0.8} binary, although the behaviour
of $\Upsilon$ is qualitatively similar for all configurations examined in
this work. More specifically, shown with a colormap is the tidal-force
indicator suitably rotated in each panel so that the green arrow always
points in the direction of the BH. The orange line indicates the region
of the NS that will be subject to mass-shedding first, while the dashed white lines
mark the rest-mass density contour where $\rho = 6.2 \times 10^{9} \,{\rm g}/{\rm{cm}^{3}}$.
Overall, the study of $\Upsilon$ highlights the growth of the tidal
forces in the centre of the NS, as well as the gradual
onset of mass shedding and the dipolar deformation that is introduced in
the direction of the BH (see green arrow).

\begin{figure}
  \centering
  \includegraphics[width=0.85\columnwidth]{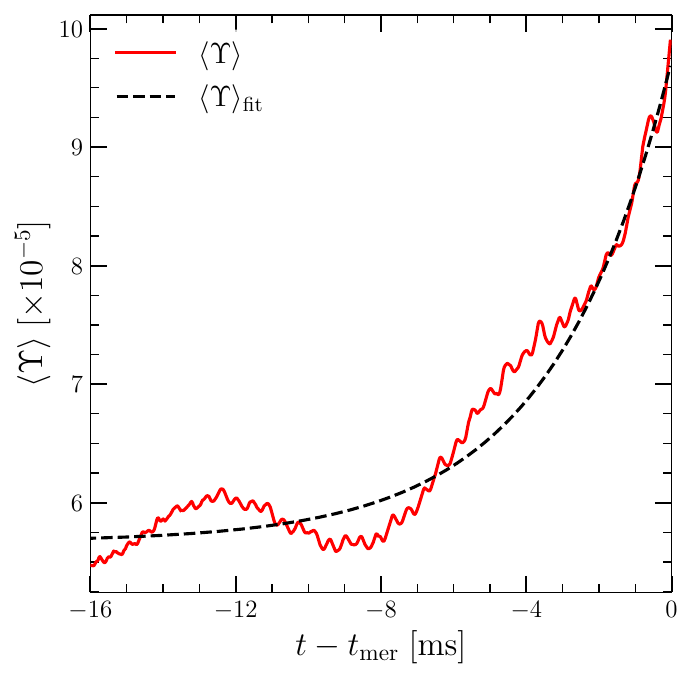}
  \caption{Evolution of the volume-averaged tidal-force indicator
    $\langle \Upsilon \rangle (t)$ during the inspiral and up to merger
    of the binary \texttt{Q7.chi0.8} (red solid line). Note how tidal
    disruption can be revealed by the onset of an exponential growth of
    $\langle \Upsilon \rangle (t)$ (black dashed line); the oscillations
    for $t-t_{\rm mer} \lesssim 8\,{\rm ms}$ are due to the eccentricity
    that has not been reduced for this binary.}
  \label{fig:evol_q7_tidalforcetimeseries}
\end{figure}

Intrigued by the spatial behaviour of the tidal-force indicator, we have
explored whether its time evolution can be used to reveal the actual
timescale for the tidal disruption. In particular, we have computed the
time evolution of the volume averaged quantity $\langle \Upsilon \rangle
(t)$ confined to the NS (\ie within a rest-mass density of $\rho = 6.2
\times 10^{9} \, {\rm g}/{\rm{cm}^{3}}$)
\begin{equation}
  \langle \Upsilon \rangle (t) := \frac{\int_{_{\rm NS}}
  \Upsilon \sqrt{\gamma}d^{3} x}{\int_{_{\rm NS}}\sqrt{\gamma}d^{3} x}\,.
  \label{eq:tdforce_avg}
\end{equation}
which is reported in Fig.~\ref{fig:evol_q7_tidalforcetimeseries} as a
solid red line. When looking at the evolution of $\langle \Upsilon
\rangle (t)$ it is not difficult to distinguish three main behaviours:
(i) an almost harmonic and low-frequency oscillation in the early stages
(\ie for $t - t_{\rm mer} \lesssim -8\,{\rm ms}$); (ii) a series of
superposed and high-frequency oscillations present at all times; (iii) a
rapid growth in the final stages (\ie for $t - t_{\rm mer} \gtrsim
-8\,{\rm ms}$). The first behaviour is clearly related to the presence of
a small, but spurious eccentricity in the initial data, while the second
one is associated with the oscillations of the star and can be observed
also in the evolution of the maximum rest-mass density. The third and
most important behaviour, instead, is signalling -- via its exponential
nature -- the presence of an instability and hence can be used to study
not only the onset of tidal disruption, but also to quantify the
characteristic timescale over which it takes place. While for compactness
we do not explore this possibility here, the disruption timescale can in
principle be associated with the physical properties of the BHNS, namely,
the EOS, the mass ratio, and the BH spin.

Hence, after removing the initial oscillating part due to eccentricity in
terms of a simple sinusoidal function $a_1 \sin \left(\omega t +
\phi\right) + a_2$, we model the evolution of $\langle \Upsilon \rangle$
as
\begin{equation}
  \langle \Upsilon \rangle_{\rm fit}(t) = \Upsilon_0
  \exp(t/\tau_{\rm disr}) \,,
  \label{eq:tdforce_ansatz}
\end{equation}
where $\Upsilon_0=4.05$ and $\tau_{\rm disr} = 3.26\,{\rm ms}$. The
corresponding fit is shown with a black dashed line in
Fig.~\ref{fig:evol_q7_tidalforcetimeseries} and clearly captures very
well the evolution of the tidal-deformation factor.

\begin{figure*}
  \centering
  \includegraphics[width=1.0\textwidth]{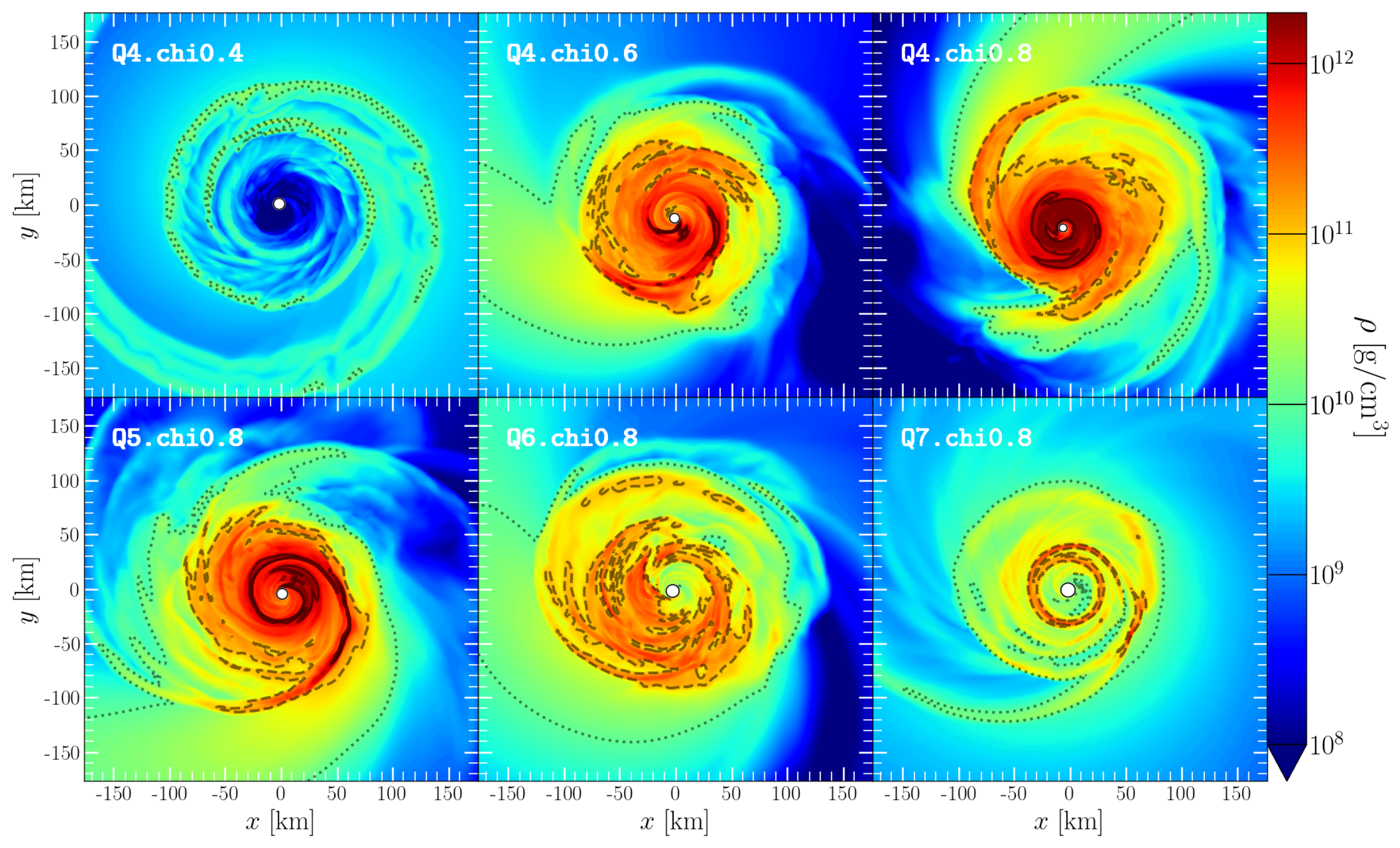}
  \caption{Snapshots at $t - t_{\rm mer} \approx 15\, {\rm ms}$ and on the orbital
    $(x,y)$ plane of the rest-mass densities for a representative subset of the
    BHNS binaries simulated in this work. The top row illustrates binaries with
    constant mass ratio ($Q=4$) and increasing BH spin from left to right
    ($\chi_{_{\rm BH}}=[0.4, 0.6, 0.8]$). The bottom row shows instead binaries
    with constant BH spin ($\chi_{_{\rm BH}}=0.8$) and varying mass ratio ($Q=[5,
    6, 7]$). Contours of the rest-mass density $\rho=10^{10}, 10^{11}, 10^{12} \,
    {\rm g}/{\rm{cm}^{3}}$ are plotted as dotted, dashed and solid black lines,
    respectively. All configurations examined here lead to a tidal disruption, but
    with considerably different disk masses and structures (see also
    Fig.~\ref{fig:postmerdis_YZ}).}
  \label{fig:postmerdis_XY}
\end{figure*}

In terms of the GW signal, the corresponding amplitude is generally
smaller as a result of the longitudinal deformation of the NS, which
reduces the quadrupole moment of the binary (see
Fig.~\ref{fig:GW_q4567_chiBHs}) and has been noted already early on in
the literature~\cite[see \eg][]{Etienne2007b,
  ShibataTaniguchi2008}. Hence, in a tidal-disruption scenario one should
expect a significant loss of power in the GW signal close to the merger
time as well as considerable quenching of the BH quasi-normal modes
(QNMs), as the BH ringdown is disturbed by the continuous and copious
accretion of matter. All of these aspects will be discussed in greater
detail in Sec.~\ref{sec:results_GW}.

In summary, in the case of the \texttt{Q4.chi0.8} binary, the tidal
disruption of the NS leads to a much larger amount of matter building up
an accretion disk around the newly formed BH, but also a smaller transfer
of angular momentum and hence spin-up (for \texttt{Q4.chi0.0} and
\texttt{Q4.chi0.8}, $\Delta \mathcal{S} = 22.26 \, M_{\odot}^{2}$ and 
$\Delta \mathcal{S}=12.51 \, M_{\odot}^{2}$, respectively).
At the same time, the GW signal is decreased in amplitude
and the ringdown quenched by the intense accretion of matter.

\section{Results: accretion disk and BH remnant}
\label{sec:results_remnant}
In this section we provide a more detailed discussion of the BHNS
simulations performed here by examining the spatial and density
distributions of the remnant disks, as well as final spins and masses of
the remnant BHs which are summarized in Tab.~\ref{tab:sim_diag}. For each
binary, the evolution is carried out until approximately $15\, {\rm ms}$
after merger, when the remnant material around the most extreme systems
has settled into a coherent disk. Evolutions on much longer timescales
and up to $\mathcal{O}(1{\rm s})$ are in principle possible (see, \eg
\cite{Kyutoku2021a}.) and can be performed with reduced computational
costs when making use of the new hybrid approach introduced in
Ref.~\cite{Ng2024b}, which will be the focus of future works.

\subsection{Accretion-disk rest-mass densities and temperatures}

\begin{figure*}
  \centering
  \includegraphics[width=1.0\textwidth]{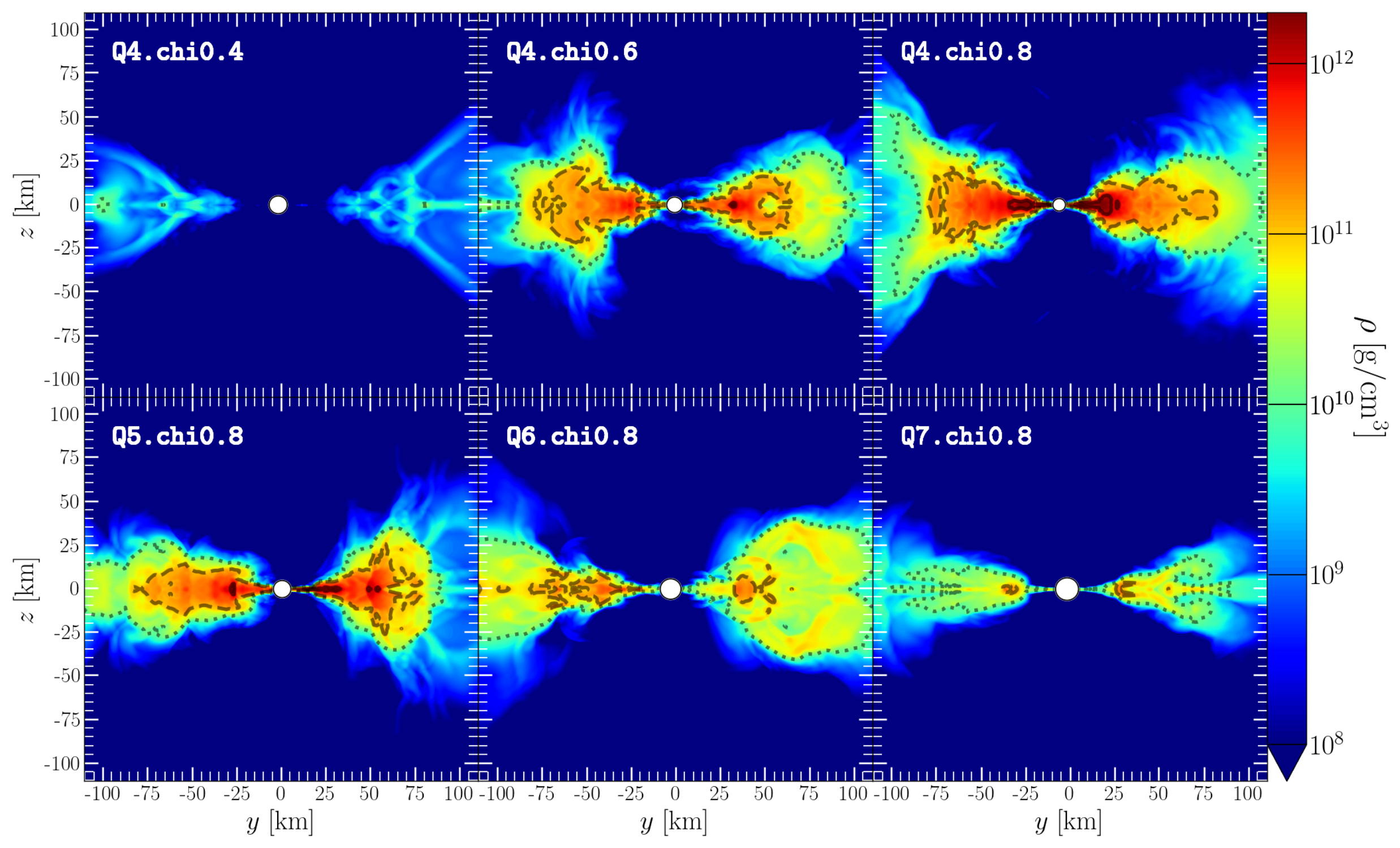}
  \caption{The same as in Fig.~\ref{fig:postmerdis_XY}, but for the
    meridional $(x,z)$ plane. Note that in the case of the binary
    \texttt{Q7.chi0.8}, there is very little matter around the BH
    immediately after the coalescence and that the disk is formed much
    later as a result of fall-back accretion.}
  \label{fig:postmerdis_YZ}
\end{figure*}

Figure~\ref{fig:postmerdis_XY} reports the distributions of the rest-mass
density for the quasi-equilibrium BHNS binary evolutions resulting in
tidal disruption of the NS where we focus on a spatial extent of
$[-175\,\rm{km}, 175\,\rm{km}]^{2}$ and at a time $t - t_{\rm mer}
\approx 15\, {\rm ms}$, which is when the remnant disks have reached a
quasi-stationary state. More specifically, the top row shows binaries
with the same mass ratio, but increasing BH spin from left to right and,
therefore, helps to identify how the BH spin helps disrupt the NS and,
hence, produce a more massive disk. As a result, we see a significant
suppression in the amount of remnant material outside of the BH for
\texttt{Q4.chi0.4} which systematically increases with increasing
prograde spin of the BH when examining \texttt{Q4.chi0.6} and
\texttt{Q4.chi0.8}~\cite{Etienne:2008re, Kyutoku2011c}.  The bottom row,
on the other hand, shows binaries with the same BH spin, but with
increasing mass asymmetry from left to right, thus highlighting how very
small mass ratios ($q \ll 1$) suppress tidal disruption and, hence,
reduce the amount of remnant material outside of the BH, which we mark
with a filled white circle\footnote{Note that the size of the BH in our
gauge is essentially given by the areal radius and the latter is
associated with the BH irreducible mass, $M_{\rm irr}$. As a result,
rapidly spinning BHs, which have smaller irreducible masses, will have a
smaller size despite having identical (gravitational) masses, $M_{\rm
  Ch}$, as is the case for binaries in the top row.}. 
Obviously, in our choice of representative binaries we have
excluded \texttt{Q4.chi0.0}, which is the pure ``plunge'' configuration
as discussed earlier, and for which the rest-mass density profile would be
below the lower bound of the colormap.

Upon closer inspection, it is easy to appreciate that the most prominent
accretion disks are found for the \texttt{Q4.chi0.8} and
\texttt{Q5.chi0.8} configurations. The rest-mass density of these disks
reaches maximum values of the order of $\rho\approx 10^{12}\,{\rm
  g}/{\rm{cm}^{3}}$, with massive tidal tails observed, which will
continually replenish the material lost to accretion during the initial
disruption. The extent and density distribution of the disks produced by
these two configurations are followed closely by the binary
\texttt{Q4.chi0.6}, where the spatial extent is slightly smaller and
lower rest-mass densities are measured throughout the disk.

When increasing the mass ratio $Q$, as in the \texttt{Q6.chi0.8} binary,
the rest-mass density is distributed much more uniformly over a similar
spatial extent, with only a small region involved in strong spiral
shocks. Similarly, in the most extreme case of the \texttt{Q7.chi0.8}
binary, the rest-mass density is considerably smaller and reaches at most
$\rho \approx 10^{11} \, {\rm g}/{\rm{cm}^{3}}$ in restricted spatial
regions. Furthermore, an increase in mass asymmetry also leads to the
tidal tail becoming systematically thinner as progressively smaller
regions of the NS are involved in the disruption process.

It is also interesting to note that axisymmetry in the disk is never
reached in our simulations, at least over the timescales we have
considered. In particular, by the simulation end time, a considerable
asymmetric inflow of matter (and angular momentum) is still present from
the tidal tail (\eg see the top edge of the top-right panel in
Fig.~\ref{fig:postmerdis_XY}). While an approximate axisymmetry will be
reached on a timescale of $\approx 30\,{\rm ms}$ after the merger, the
trace of the tidal tail will be present for longer times (see also
\cite{Kyutoku2015}).

Complementary information on the structure of the remnant disk is offered
in Fig.~\ref{fig:postmerdis_YZ}, which reports the rest-mass density
distributions in the $(x,z)$ plane for the same binaries discussed in
Fig.~\ref{fig:postmerdis_XY}. Overall, the qualitative behaviour in the
meridional direction is similar to that encountered in the equatorial one
and both the spatial extent and the height become systematically
suppressed as the mass asymmetry increases [see also the discussion in
  Ref.~\cite{Etienne:2008re}]. In addition, the \texttt{Q4.chi0.8} binary
leads to the accretion disk with the largest vertical extent and the most
extended high-density regions (see the dashed contour with $\rho =
10^{11} \, {\rm g}/{\rm{cm}^{3}}$), while the \texttt{Q4.chi0.4} binary
produces only a very tenuous disk structure.

\begin{figure*}
  \centering
  \includegraphics[width=1.00\textwidth]{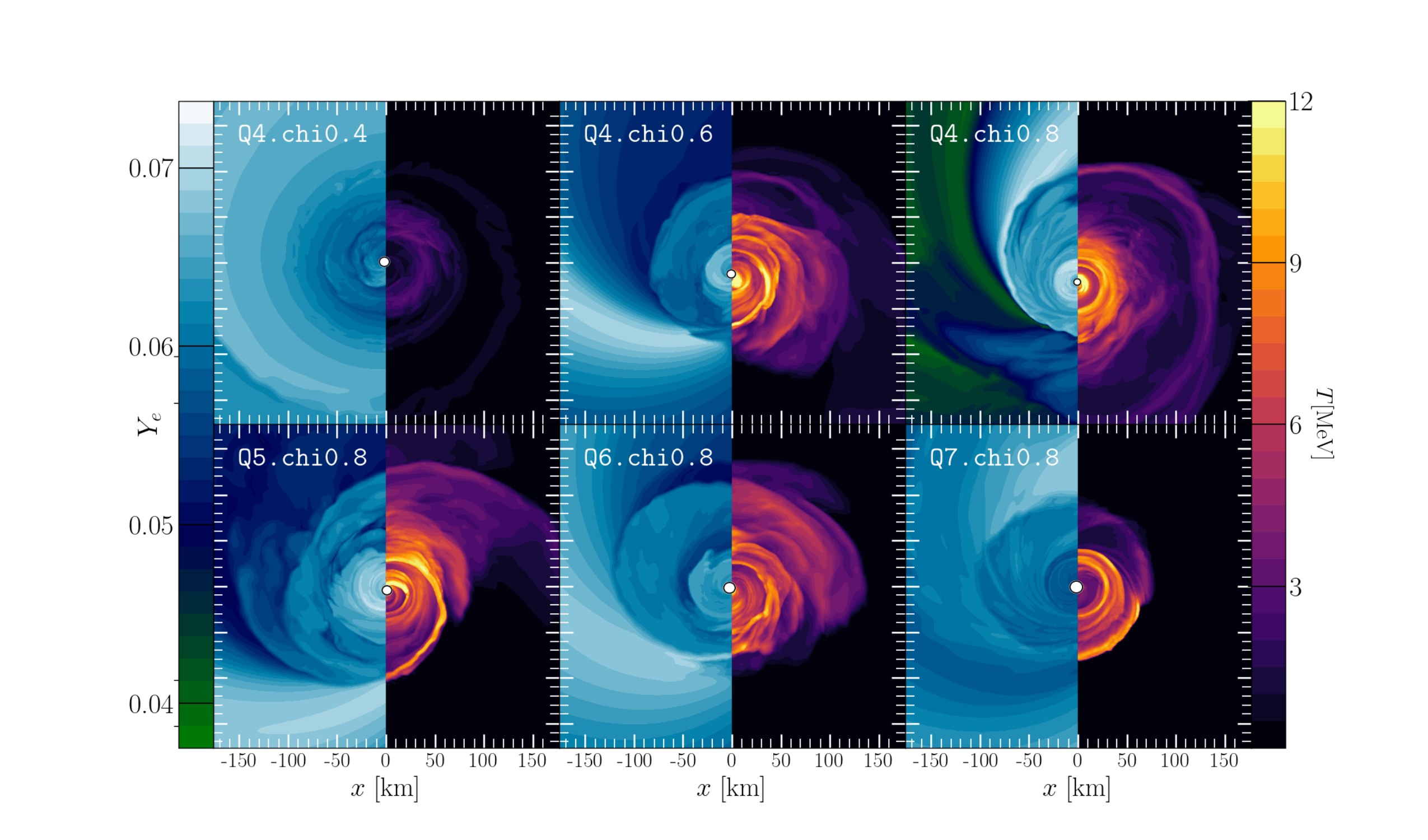}
  \caption{The same as in Fig.~\ref{fig:postmerdis_XY}, but for the
    temperature (right part of each panel) and electron-fraction (left
    part of each panel) distributions.}
  \label{fig:postmerdis_XY_temp}
\end{figure*}

An important observation to make is that, at least for the EOS considered
here, the overall structure of the remnant accretion disk changes
considerably between the mass ratios $Q=5$ and $Q=6$, thus separating the
space of parameters where tidal disruption produces visually distinct
outcomes and leaves behind matter with significantly different rest-mass
densities. More specifically, the inspection of
Fig.~\ref{fig:postmerdis_XY} shows that accretion disks for $Q>5$ do not
feature a region of very high density material, \ie where $\rho > 10^{12}
\, {\rm g} / {\rm{cm}^{3}}$. This is best appreciated by following the
solid contour of corresponding density, which does not appear for
$Q>6$. Similarly, matter enclosed within the $\rho = 10^{11} \, {\rm g} / {\rm{cm}^{3}}$ 
contour is scarcely present in
the meridional slice of the accretion disk in
Fig.~\ref{fig:postmerdis_YZ} for the $Q=6$ case, and disappears almost
completely for $Q=7$. Simultaneously, the polar extent of the disk
changes substantially between $Q=5$ and $Q=6$. Last but not least, by
inspecting the dotted contour related to $\rho = 10^{10} \, {\rm
  g}/{\rm{cm}^{3}}$, we see that the addition of prograde spin angular
momentum of the BH increases the opening angle and polar distribution of
the remnant material, as seen in the binaries \texttt{Q4.chi0.4},
\texttt{Q4.chi0.6}, and \texttt{Q4.chi0.8}. Similarly, for fixed BH spin
and decreasing mass ratio $q$, as seen in the binaries \texttt{Q5.chi0.8},
\texttt{Q6.chi0.8}, and \texttt{Q7.chi0.8}, the remnant disk structure
tends to be flatter with the binaries having mass ratios $Q = 6, 7$
marking, again, the transition between vertically extended and vertically
slim remnant disks (see also discussion in Refs.~\cite{Foucart2013a,
  Foucart2014}).

Finally, although the timescale covered by our simulations is not large
and hence cannot cover secular effects such as the heating generated by
MHD-driven turbulent viscosity or neutrino cooling/heating, we briefly
discuss the temperature properties of the remnant disks discussed above
right after merger, when the highest temperatures are expected to be
reached. In particular, we complement the information on the rest-mass
density provided with Fig.~\ref{fig:postmerdis_XY} at time $t-t_{\rm mer}
\approx 15\,{\rm ms}$, by reporting in Fig.~\ref{fig:postmerdis_XY_temp}
the corresponding temperature distributions, respectively in the right
half of each panel. The various panels are rather self-explanatory and
clearly indicate that binaries with large mass asymmetries and smaller BH
spins not only lead to smaller disks, but also that they are ''cold'',
with maximum temperatures $T_{\rm max} \lesssim 10 \, {\rm MeV}$ and
average temperatures $T_{\rm avg} \approx 6 \, {\rm MeV}$ (an even colder
disk is present for the binary \texttt{Q4.chi0.4} on the top-left
panel). When keeping the mass ratio the same but increasing the BH spin
(see binaries \texttt{Q4.chi0.6} and \texttt{Q4.chi0.8} on the top-middle
and top-right panels), both maximal and average temperatures in the disk
reach larger values $T_{\rm max} \approx 15 \, {\rm MeV}$ and
$T_{\rm avg} \approx 10 \, {\rm MeV}$. These values, which are in very
good agreement with those found in Ref.~\cite{Most2021a} when employing
two different EOSs, are still relatively cold when compared with the
temperatures found in BNS mergers which can be a factor 4-5 larger.

On the other hand, the temperature behaviour has less variance when
considering binaries with the same large BH spin and increasing degree of
mass asymmetry. In this case, in fact, (see binaries \texttt{Q5.chi0.8},
\texttt{Q6.chi0.8}, and \texttt{Q7.chi0.8} on the bottom panels of
Fig.~\ref{fig:postmerdis_XY_temp}) both the maximum and average
temperatures do not show significant changes across the different
binaries and reach $T_{\rm max} \approx 15\, {\rm MeV}$, and average
temperatures $T_{\rm avg} \approx 10~\rm MeV$, although the binary with
$Q=7$ has somewhat lower temperatures. This behaviour is the result of
the dynamics of the tidal disruption and shock heating resulting from the
collision of disrupted material and of ejected material that rapidly
falls back onto the remnant. Under these conditions, strong shocks are
inevitable and lead to large local and global temperatures. Hence,
so long as a tidal disruption takes place, the associated disk
temperatures are expected to be large, with the maximum values being
directly associated with the rapidity and severity of the
disruption. This also explains why the binary \texttt{Q7.chi0.8}, which
experiences a comparatively milder disruption, will also have
comparatively colder accretion disk.

For completeness, we also report on the left part of each panel in
Fig.~\ref{fig:postmerdis_XY_temp} the distributions of the electron
fraction $Y_{e}$. Although we are not employing here any
radiative-transfer treatment for the neutrinos, such distributions
provide a reasonable description thanks to the very short timescales
involved with the tidal disruption and merger. The crucial observation is
that a strong tidal disruption promotes the presence of neutron-rich
material with lower $Y_{e}$ at sites further away from the accretion
disc, and also visible in the traces of the tidal tail still connected to
the disc. Indeed, the \texttt{Q4.chi0.8} and \texttt{Q5.chi0.8} binaries
feature the lowest $Y_{e}$, with dark-blue and green regions
corresponding to the $Y_{e}\sim 0.04-0.05$; this is especially visible
for the \texttt{Q4.chi0.8} binary, for which a spiral-like structure in
the upper left corner is visibly neutron-rich and reminiscent of the
prior tidal disruption dynamics. On the other hand, less strong
disruptions, \eg binaries \texttt{Q4.chi0.4}, \texttt{Q4.chi0.6}, and
\texttt{Q6.chi0.8}, display a slightly higher level of $Y_{e}\approx
0.06$ for the outer parts of the disc. This is because a strong
disruption initially preserves a large amount of neutron-rich material
from the neutron star.

However, when considering the innermost part of the accretion disk, a
different trend is observed. More specifically, when considering a
coordinate radius of $r=25\,{\rm km}$, the binaries \texttt{Q4.chi0.8}
and \texttt{Q5.chi0.8}, which have undergone a strong disruption, exhibit
material of $Y_{e} \gtrsim 0.07$, visible as white regions around the
apparent horizon. On the other hand, for their less disruptive
counterparts \texttt{Q6.chi0.8} and \texttt{Q7.chi0.8}, we find
$Y_{e}\approx 0.065$ and $Y_{e}\approx 0.055$, respectively. A comparison
with the adjoined right panels in Fig.~\ref{fig:postmerdis_XY_temp}
proves that this behaviour is tightly correlated with temperatures
reached in the innermost part of the disk and hence $Y_{e}$ in this
region is driven primarily by the local thermodynamical equilibrium.

In summary, the extensive series of simulations performed here has
highlighted that large mass-asymmetries or reduced BH-spins will decrease
the amount of matter composing the remnant disk, lowering the
corresponding rest-mass densities and temperatures. By contrast, small
mass-asymmetries, but large prograde BH spins will lead to increasingly
more massive disks with significantly higher rest-mass densities and
temperatures on average. In all cases, the radial extent of the remnant
disks in the equatorial plane is $\simeq 100\,{\rm km}$, while the
vertical extension can vary by a factor of two, between $\simeq 50\,{\rm
  km}$ and $\simeq 25\,{\rm km}$. These results are qualitatively
consistent with those already presented in Refs.~\cite{Etienne:2008re,
  Foucart2010, Kyutoku2011}.

\subsection{Accretion-disk masses}
\label{sec:mdisk}

\begin{figure}
  \centering
  \includegraphics[width=0.94\columnwidth]{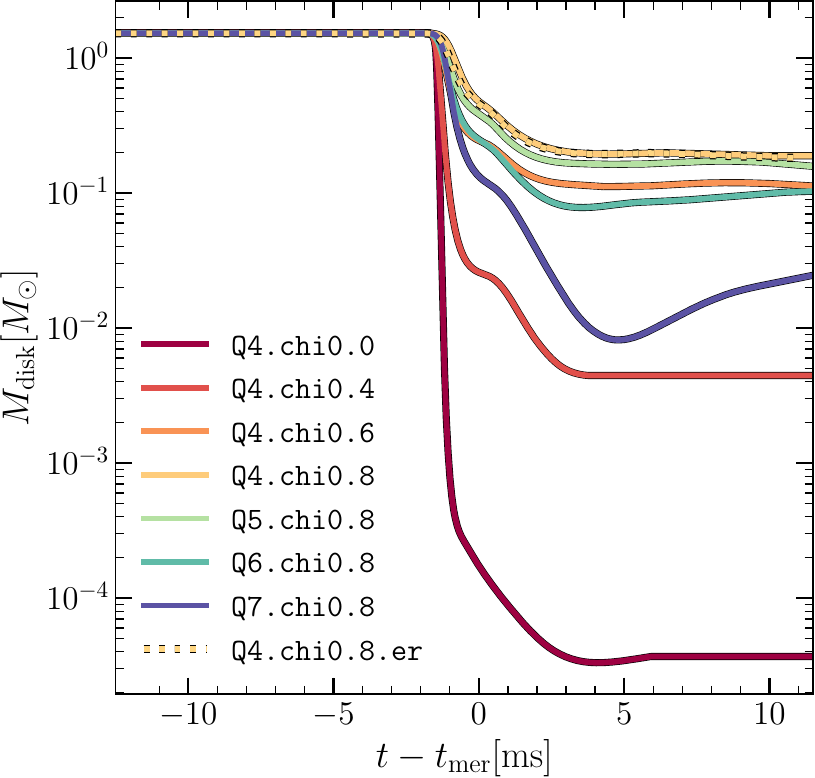}
  \caption{Evolution of the rest-mass of the accretion disk $M_{\rm
      disk}$ for all BHNS binaries simulated here. Note how the disk mass
    can easily be used to mark the transition from a plunge to a
    disruption scenario; for the EOS considered here, this happens for
    the binary \texttt{Q4.chi0.0} for which $M_{\rm disk}\leq 10^{-2} \,
	M_{\odot}$.}
  \label{fig:BHNS_rem_mass}
\end{figure}

We next turn to a more quantitative discussion of the amount of matter
composing the remnant disk by computing $M_{\rm disk}$ as the integral of
the rest-mass density [see Eq.~\eqref{eq:disc_mass_int}]. The integration
volume is set by a coordinate radius of $r_{\rm disk} \approx 100\, {\rm
  km}$ as this provides a robust estimate of the material actively
participating in the disk dynamics. The results are collected in
Fig.~\ref{fig:BHNS_rem_mass}, which reports the evolution of the remnant
disk mass in terms of the retarded time since the merger, \ie $t-t_{\rm
  mer}$. In this way it is straightforward to appreciate the three most
salient aspects of this quantity.

First, it undergoes a significant reduction well before $t-t_{\rm mer} =
0$ as the latter marks the time when the GW amplitude reaches its first
maximum and this actually takes place after part of the NS mass has been
accreted by the BH. Second, for a given mass ratio, the spin plays a
fundamental role in determining the remnant-disk mass, which can vary by
more than four orders of magnitude. In particular, because a small spin
will favour a plunge, the remnant disk is the smallest for the binary
\texttt{Q4.chi0.0}, while it is the largest and
$\mathcal{O}(0.1\,M_{\odot})$ for the \texttt{Q4.chi0.8} binary, with
intermediate spins filling-in between these two extremes in a nonlinear
manner (see Tab.~\ref{tab:sim_diag} for details). Third, the evolution of
$M_{\rm disk}$ is not monotonic in time and indeed all binaries exhibit
the presence of a local minimum after the substantial prompt accretion
phase. This is due to the fallback of ejected yet bound material, which
is clearly visible even over the short timescales investigated here. As
the mass asymmetry $Q$ increases, more material is initially ejected
outside of the integration region and to larger spatial extents requiring
longer timescales for the matter to fallback into the integration
domain. This behaviour can be tracked via the local minimum in $M_{\rm
  disk}$, that can easily be identified for the \texttt{Q6.chi0.8}
binary, and which systematically occurs at later times and becoming very
pronounced for the binary \texttt{Q7.chi0.8}. Finally, the reduction of
eccentricity, which has been applied to the \texttt{Q4.chi0.8} binary
(\ie case \texttt{Q4.chi0.8.er}) hardly modifies the properties of the
remnant disk, hence underlining that eccentricity reduction is most
important in modelling the gravitational signal and, in particular, its
ringdown part (see discussion in Sec.~\ref{sec:gw_ringdown}).

\begin{table*}[t]
  \begin{ruledtabular}
    \begin{tabular}{c|crccccccr|cc}
      binary & $\chi_{_{\rm BH}}^{\rm rem} $ & $M_{\rm Ch}^{\rm rem}$ &
      $M^{\rm rem}_{\rm irr}$ & $M_{\rm disk}$ & $M_{\rm ej}$ & $M_{\rm b, rem}$ & $\hat{M}_{\rm rem}$ &
      $f_{\rm mer}$ & $\tau_{_{\rm sur}}$ & $f_{_{\rm MS}}$ & $f_{_{\rm ISCO}}$ \\
      & & $[M_{\odot}]$ & $[M_{\odot}]$ & $[M_{\odot}]$ & $[10^{-1}
        M_{\odot}]$ & $[M_{\odot}]$ &
      & $[$kHz$]$ & [$\rm ms$] & $[$kHz$]$ & $[$kHz$]$ \\

      \hline
      \texttt{Q4.chi0.0~~~} & $0.474$ &  $6.853$ & $6.645$ & $\approx 0$ & $\approx 0$ & $\approx 0$ & $\approx 0 $ & $1.381$ & $-0.1$ & $0.919$ & $0.889$ \\
      \texttt{Q4.chi0.4~~~} & $0.679$ &  $6.836$ & $6.365$ & $0.004$     & $ 0.063 $   & $ 0.024$    & $0.016$      & $1.386$ & $0.7$  & $0.930$ & $1.201$ \\
      \texttt{Q4.chi0.6~~~} & $0.772$ &  $6.658$ & $6.021$ & $0.114$     & $ 0.195 $   & $ 0.202$    & $0.132$      & $1.385$ & $1.1$  & $0.933$ & $1.458$ \\
      \texttt{Q4.chi0.8~~~} & $0.868$ &  $6.582$ & $5.694$ & $0.189$     & $ 0.427 $   & $ 0.293$    & $0.192$      & $1.282$ & $1.9$  & $0.935$ & $1.783$ \\
      \texttt{Q5.chi0.8~~~} & $0.861$ &  $7.979$ & $6.930$ & $0.160$     & $ 0.414 $   & $ 0.275$    & $0.180$      & $1.284$ & $1.6$  & $0.893$ & $1.415$ \\
      \texttt{Q6.chi0.8~~~} & $0.858$ &  $9.414$ & $8.190$ & $0.103$     & $ 0.282 $   & $ 0.212$    & $0.138$      & $1.242$ & $1.1$  & $0.858$ & $1.171$ \\
      \texttt{Q7.chi0.8~~~} & $0.858$ & $10.890$ & $9.474$ & $0.025$     & $ 0.233 $   & $ 0.108$    & $0.070$      & $1.162$ & $0.7$  & $0.826$ & $0.998$ \\
      \hline
      \texttt{Q4.chi0.8.er} & $0.868$ & $6.582$ & $5.694$ & $0.189$     & $ 0.427 $   & $ 0.293$     & $0.192$      & $1.211$  & $1.9$ & $0.935$ & $1.783$ \\
    \end{tabular}
  \end{ruledtabular}
  \caption{Diagnostic information about the simulations performed in this
    work together with the corresponding QE predictions. Reported are:
    the remnant BH spin $\chi_{\rm BH}^{\rm rem}$; its gravitational mass
    $M_{\rm Ch}^{\rm rem}$ and irreducible mass $M_{\rm irr}^{\rm rem}$;
    the accretion-disk mass $M_{\rm disk}$; the ejected mass $M_{\rm
      ej}$; the remnant baryonic mass $M_{\rm b, rem}$ and its value
    normalized by the NSs initial baryonic mass $\hat{M}_{\rm b, rem}$;
    the frequency of the $\ell=2=m$ mode of the GW at merger $f_{\rm
      mer}$; and the survival time $\tau_{\rm sur}$ of the stripped
    matter before accretion. Finally, also reported are the mass-shedding
	and ISCO frequencies, $f_{_{\rm MS}}$ and $f_{_{\rm ISCO}}$
    respectively, derived from QE sequences for the DD2 EOS, as discussed
    in Appendix~\ref{sec:qe_sequences}.}
\label{tab:sim_diag}
\end{table*}

It should be noted that the disk mass [as defined in
  Eq.~\eqref{eq:disc_mass_int}] is smaller than the bound rest-mass
$M_{\rm bound}$ [which is instead introduced in
  Eq.~\eqref{eq:disc_mass_bound}]. This is because $M_{\rm bound}$ is
computed using a larger integration domain than for $M_{\rm disk}$ and,
as commented above, encompasses the bound rest-mass in the
proto-accretion disk in addition to the bound matter that has been
ejected and will fallback over longer timescales than those considered
here. The values of $M_{\rm disk}$, $M_{\rm b, mer}$, of the normalized
remnant mass $\hat{M}_{\rm mer}$, and of the amount of unbound matter
$M_{\rm ej}$ as computed at the end of each simulation are reported in
Tab.~\ref{tab:sim_diag} and are measured before any material leaves the
computational domain.

It should be noted that for binaries with $Q\leq 6$, the ratio between
the mass in the disk and the bound material is already quite large, \ie
$M_{\rm disk}/M_{\rm bound} \gtrsim 0.5$ at times $t-t_{\rm mer} \approx
10\, {\rm ms}$. However, the amount of bound material at large distances
(\ie at $r \gtrsim 100\, M_{\odot}$) becomes even greater for larger mass
asymmetry. This is especially pronounced for the $Q=7$ simulation, where
the value of $M_{\rm disk}$ is approximately $30\%$ of $M_{\rm bound}$
at a time $t-t_{\rm mer} \approx 10\, {\rm ms}$,
which naturally follows the minimum of the curve in
Fig.~\ref{fig:BHNS_rem_mass}. Clearly, given enough time, $M_{\rm disk}$
for all configurations will approach $M_{\rm bound}$ due to the fallback
of the bound material from $r>100\,M_{\odot}$. Hence, from the magnitude
of $M_{\rm bound}$ and $M_{\rm disk}$, as well as from the slope of
$M_{\rm disk}$ at times $t - t_{\rm mer} \gtrsim 10 \,{\rm ms}$ in
Fig.~\ref{fig:BHNS_rem_mass}, we infer that BHNS mergers with $Q\geq 7$
do not contain in the remnant disk most of the bound material and that
the disk will gain a substantial amount of rest-mass over a timescale
of several tens of milliseconds. In the case of the \texttt{Q7.chi0.8}
binary, this fallback material is abundant enough to increase the mass
in the disk by a factor of two by the end of the simulation. Hence, the
disk mass would continue to increase until the accretion rate onto the BH
dominates over the rate of the fallback of the bound material onto the
disk. As discussed in Ref.~\cite{Musolino2024} for the case of BNS
binaries, this finding reveals the significant role that fallback
material can have on modelling the EM emission from these binaries. This
is particularly important for the choice of realistic initial data, since
$Q\approx 7$ is thought to be the peak of the mass ratio distribution in
the BHNS population~\cite{Broekgaarden2021}. Hence, a delayed formation
of a massive accretion disk is likely to impact the MHD processes of jet
launching, and consequently, the association of these events with GRBs.

\subsection{Total remnant rest-mass}

The analysis of the bound and disk-related parts of the residual
rest-mass naturally leads to the more general question about the total
mass left after the coalescence and over a timescale of
$\mathcal{O}(10\,\rm ms)$. In this respect, and as expected, the
\texttt{Q4.chi0.8} configuration leads to the greatest value of $M_{\rm
  b, rem} \simeq 0.3\, M_{\odot}$, while BH spins of $\chi_{_{\rm
    BH}}=0.6$ and $\chi_{_{\rm BH}}=0.4$ decrease this value to $\simeq
0.2\, M_{\odot}$ and $\simeq 0.030\,M_{\odot}$, which are $30\%$ and an
order of magnitude smaller, respectively. On the other hand, an increase
in the mass of the BH at constant BH spin yields a more moderate decrease
in $M_{\rm b, rem}$, leading to total remnant masses of $0.275\,
M_{\odot}$, $0.2\, M_{\odot}$ and $0.1\, M_{\odot}$ for the $Q=5,6,7$
configurations, respectively. We note that a degree of degeneracy is
present for the \texttt{Q4.chi0.6} and \texttt{Q6.chi0.8} binaries as the
difference in the total remnant mass is only $0.01\, M_{\odot}$, but the
ejected mass is $50\%$ larger for the \texttt{Q6.chi0.8}
configuration. This implies that spin and mass-ratio inference from a
post-merger EM signal could be facilitated by a detailed modelling of the
nucleosynthetic yields and subsequent kilonova emission (see paper III
for a discussion).

Our measurement of the total remnant mass can be straightforwardly
compared with the predictions suggested in Ref.~\cite{Foucart2018b}, where the
remnant mass is modeled via an analytic expression combining a large
number of BHNS simulations. In such a model, which we denote as
$\hat{M}_{\rm b, rem}^{F18}$, the predicted total remnant baryon mass
normalized by the baryon mass of the NS is parametrized in terms of
stellar compactness $\mathcal{C}$, symmetric mass ratio $\eta$, and BH
dimensionless spin $\chi_{_{\rm BH}}$, reading~\cite{Foucart2018b}
\begin{align}
  \hat{M}_{\rm rem}^{\rm{F}18} :=\left[
    {\rm{Max}} \left\lbrace \alpha \left( 1-2\mathcal{C} \right) \eta^{-1/3} - \beta
    \hat{R}_{\rm ISCO} + \gamma \right\rbrace, 0 \right]^{\delta}\,, \nonumber
\end{align}
where $\eta := Q/(1+Q)^{2}$ is the symmetric mass ratio, and
$\hat{R}_{\rm ISCO}$ is the radial coordinate of the ISCO in Kerr
spacetime for Boyer-Lindquist coordinates, rescaled by the BH mass. The
parameters $(\alpha,\beta,\gamma,\delta)$ have best-fit values
$\alpha=0.406$, $\beta=0.139$, $\gamma=0.255$, $\delta=1.761$.

When comparing our results to the predictions of this model, we find a
relative error $\lesssim 30\%$ for most of the simulations. More
specifically, for the \texttt{Q7.chi0.8} binary the relative difference
in $\hat{M}_{\rm b, rem}$ and $\hat{M}_{\rm b, rem}^{F18}$ is $40\%$,
with our numerical result being smaller than the prediction of the
fit. For the \texttt{Q4.chi0.4} binary, on the other hand, the relative
difference reaches $56\%$, again with the fit overestimating the measured
results. While these differences may appear large, we should recall that
the differences found between measurements and the fit predictions
$\mathcal{O}(50\%)$ are well within the uncertainties in the estimates of
$\hat{M}_{\rm b, rem}^{F18}$~\cite{Foucart2018b}.

In addition to a comparison with the predictions from the fit of
Ref.~\cite{Foucart2018b}, we can also perform a direct comparison with
other published BHNS simulations which employ initial data that is similar to
the one considered here. In particular, we can compare the estimates from
the binaries \texttt{Q4.chi0.8} and \texttt{Q6.chi0.8} with similar
binaries presented in Ref.~\cite{Hayashi2023}, where the EOS is the same,
the NS has a mass of $M_{_{\rm NS}}=1.35\, M_{\odot}$, and the BH a spin
$\chi_{_{\rm BH}}=0.75$. While these values differ by about $5\%$ from
ours, we expect the discrepancies to compensate each other to some
degree, since a $3\%$ higher compactness for our NS
is compensated by the slightly larger spin of our BH.

Comparing the measured value of $M_{\rm b, rem}$ in
Ref.~\cite{Hayashi2023} at $t - t_{\rm mer}\approx 20 \, {\rm ms}$, \ie
before matter accretion onto the black hole starts to be driven primarily
by angular momentum transport due to magnetohydrodynamically induced
viscosity \cite{BalbusHawley1998}, we find a difference of only $3\%$ and
$4\%$ for the $Q = 4$ and $Q=6$ binaries, respectively. This match is
quite striking despite the many potential factors involved, \ie the
differences in BH spin and NS mass, the differences in the numerical grid
structure, the grid resolutions used and the order of accuracy of
numerical methods utilized. Similar precisions are found when comparing
the ejected masses, which are $M_{\rm ej}\approx 0.045\, M_{\odot}$
($M_{\rm ej}\approx 0.035\, M_{\odot}$) for the $Q=4$ ($Q=6$)
binary~\cite{Hayashi2023}, to be contrasted with our measurement of
$M_{\rm ej}=0.042\, M_{\odot}$ ($M_{\rm ej}=0.028\, M_{\odot}$). While
larger (but still below $10\%$) these differences are also surprisingly
good given that the measurement of rest-masses associated with small
rest-mass densities remains challenging in these simulations. All things
considered, the fact that our measurements have a percent precision when
compared with independent numerical simulations suggests that the
differences highlighted above with respect to the predictions of
Ref.~\cite{Foucart2018b} are likely due to the inevitable imprecision of
the very general fitting expression for $\hat{M}_{\rm mer}^{F18}$.

\begin{figure}
  \centering
  \includegraphics[width=0.98\columnwidth]{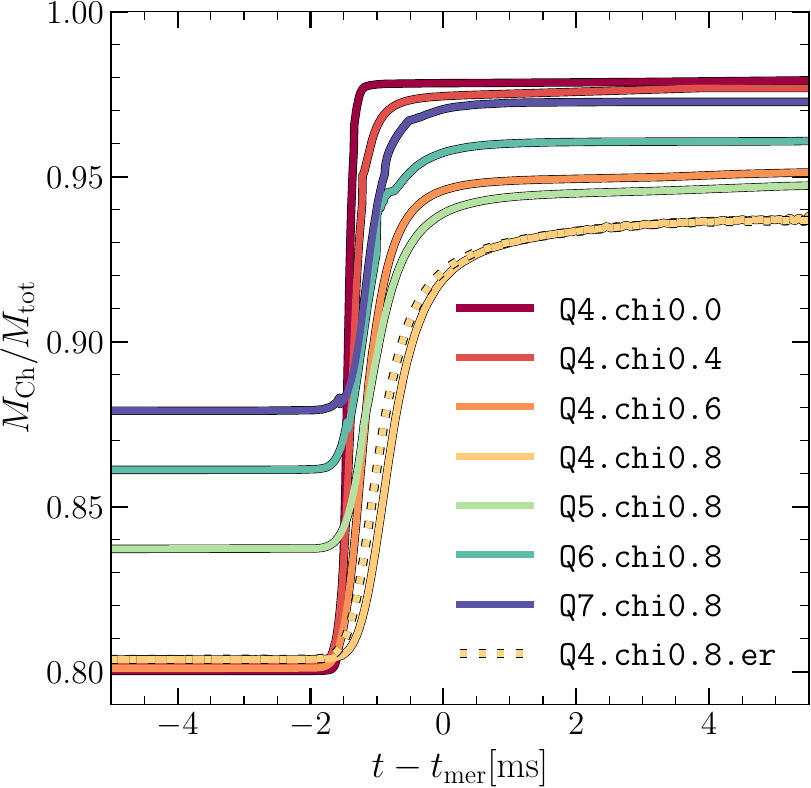}
  \caption{Evolution of the Christodoulou mass of the BH $M_{\rm Ch}$
    scaled by the total mass of the system $M_{\rm tot}$ for all BHNS
    binaries simulated here.}
  \label{fig:BHNS_sim_BH_Chmass}
\end{figure}
%

\subsection{Mass and spin of the BH remnant}
\label{subsec:results_BH_remnant}

The last topic we will cover in this Section is dedicated to the
properties of remnant BHs. We recall that for all binaries examined in
this work, the secondary (\ie the NS) is irrotational and the BH spin is
aligned with the orbital angular momentum. Under these conditions, the
generic expectation is that the BH remnant will increase its mass and
dimensionless spin after the merger and what needs to be established is
how large these changes are for different mass ratios
and initial BH spins in our binaries and how rapidly the new BH
attains its new asymptotic properties.

We start by considering the evolution of the total (Christodoulou) mass
of the BH, $M_{\rm Ch}$, for our eight BHNS configurations, which is
shown in Fig.~\ref{fig:BHNS_sim_BH_Chmass}, and where it is normalized by
$M_{\rm tot}$. Defining $M^{\rm rem}_{\rm Ch}/M_{\rm tot}$ as the
asymptotic values of $M_{\rm Ch}/M_{\rm tot}$, the corresponding
values are, in principle, dictated by the combination of the amount of mass outside
the BH after tidal disruption (which is favoured by small mass
asymmetries) and of the amount of mass lost via GW emission (which is
boosted by large BH spins). In practice, however, a rapid inspection of
Fig.~\ref{fig:BHNS_sim_BH_Chmass} reveals that it is really the degree of
tidal disruption or lack thereof that dominates the BH-mass
growth. Indeed, the largest relative growth of the BH is obtained with
the \texttt{Q4.chi0.0} binary, where the absence of a tidal disruption
and subsequent plunge of the NS into the BH leads to a substantial
increase in the mass of the BH relative to its initial mass. A similar
behaviour is consistently seen for all of the $Q=4$ configurations, where
the relative increase in BH mass (\ie $1-M^{\rm rem}_{\rm Ch}/M_{\rm
Ch}$) systematically decreases as the spin increases as a result of more
prominent tidal disruption of the NS.

It is possible to compare our values of $M^{\rm rem}_{\rm Ch}$ to the
analytical fitting model in Ref.~\cite{Zappa2019, Gonzalez2022a}, which
corrects the BH remnant model for BBH mergers built over the
years~\cite{Rezzolla-etal-2007, Rezzolla-etal-2007b, Barausse:2009uz,
  Healy2014, Hofmann2016, JimenezForteza2016} through a functional
dependence on $(M_{\rm Ch}, M_{_{\rm NS}}, \chi_{_{\rm BH}}, \Lambda)$ (see also
\cite{Pannarale2012, Pannarale2013b} for earlier work on modelling the BH
remnants of BHNS mergers). Overall, we found relative differences of less
than $1\%$ for all the configurations, except for the \texttt{Q7.chi0.8}
binary, for which the error increases to $2.5\%$. To highlight this
limitation further, we note that our \texttt{Q7.chi0.8} configuration
with $M^{\rm rem}_{\rm Ch}/M_{\rm Ch} \approx 1.11$ has very similar
parameters to M14-10-S8 in Ref.~\cite{Foucart2014}, for which likewise
$M^{\rm rem}_{\rm Ch}/M_{\rm Ch}\approx 1.1$. While this comparison
clearly indicates the fidelity of the fitting model, which is similar to
those obtained for BBH binaries~\cite{Healy2014, Hofmann2016,
  JimenezForteza2016}, the fact that the largest error is obtained for
binaries with the largest mass asymmetry also points out the need of
additional simulations with $Q=\mathcal{O}(10)$ in order to obtain a
faithful prediction of the BH mass from BHNS mergers in the whole space
of allowed parameters.

As a final remark in this context, it is worth highlighting that the
time required to reach the asymptotic value $M^{\rm rem}_{\rm Ch}$
depends on the properties of the binary and, in particular, on the
initial spin angular momentum of the BH. More specifically, while the
\texttt{Q4.chi0.0} binary experiences a very sharp growth of the BH mass
over a timescale $\lesssim 0.5\, {\rm ms}$, the mass of the BH in the
\texttt{Q4.chi0.8} binary experiences a much slower growth, which takes
place over a timescale of $\approx 4\,{\rm ms}$. This is obviously due to
the fact that for such a binary, the part of the material below the
effective ISCO has taken part in the tidal disruption process and gained
sufficient angular momentum to avoid direct accretion onto the black
hole, but not sufficiently enough to maintain a stable
orbit at or outside the ISCO. 

\begin{figure}
  \centering
  \includegraphics[width=0.94\columnwidth]{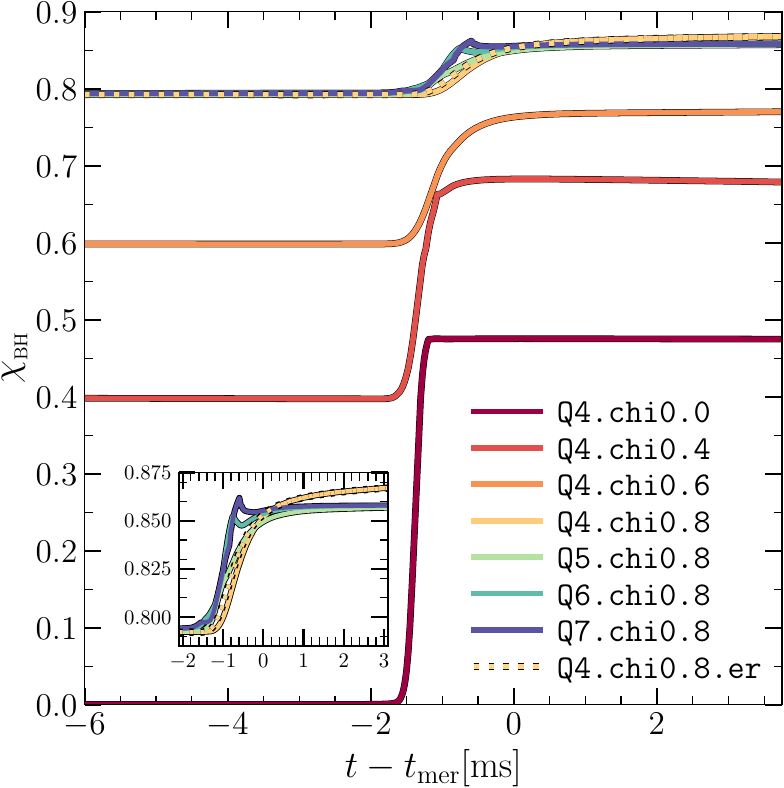}
  \caption{Evolution of the dimensionless BH spin $\chi_{_{\rm BH}}$ for all
    of the binaries simulated in this work. Note that in the binaries
    with initial BH spin $\chi_{_{\rm BH}} = 0.8$, the spin-up of the BH
    post-merger is essentially insensitive to the mass ratio. Note also
    that the rapid changes in spin reported for the binaries with $Q = 6,
    7$ at $t-t_{\rm mer}\approx -1\, {\rm ms}$ are the result of rapid
    changes of the apparent horizon coordinate surface and do not have a
    physical meaning.}
  \label{fig:BHNS_BHspin}
\end{figure}

Next, we consider the evolution of the remnant BH spin $\chi^{\rm
  rem}_{_{\rm BH}}$ for our eight BHNS configurations, which is shown in
Fig.~\ref{fig:BHNS_BHspin} and that clearly shows how, quite generically,
the orbital angular momentum of the system in the late inspiral is
transferred through matter inflow into the BH, increasing its spin. The
spin-up is therefore maximal for the initially irrotational BHNS binary
\texttt{Q4.chi0.0}, which is spun up to a final dimensionless spin of
$\chi^{\rm rem}_{_{\rm BH}}=0.474$. At the same time, the increase will
delicately depend on the interplay between the mass ratio and the
initial spin. For instance, considering binaries that have a fixed mass
ratio of $Q=4$ and contrasting the different initial BH spins
$\chi_{_{\rm BH}} = [0.4,0.6,0.8 ]$ with the corresponding final spins
$\chi^{\rm rem}_{_{\rm BH}} = [0.679, 0.772, 0.868]$, it is clear that
the higher $\chi_{_{\rm BH}}$ is, the harder it becomes to increase it
further.

\begin{figure*}
  \centering
  \includegraphics[width=0.495\textwidth]{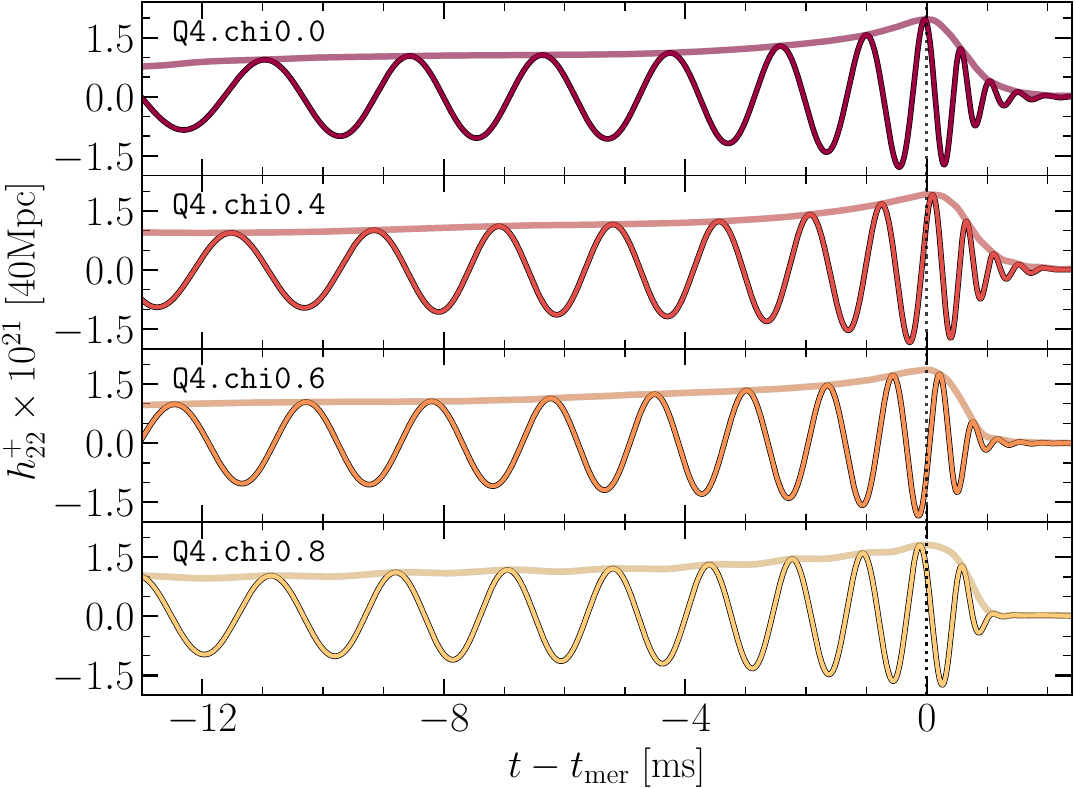}
  \hskip 0.1cm
  \includegraphics[width=0.495\textwidth]{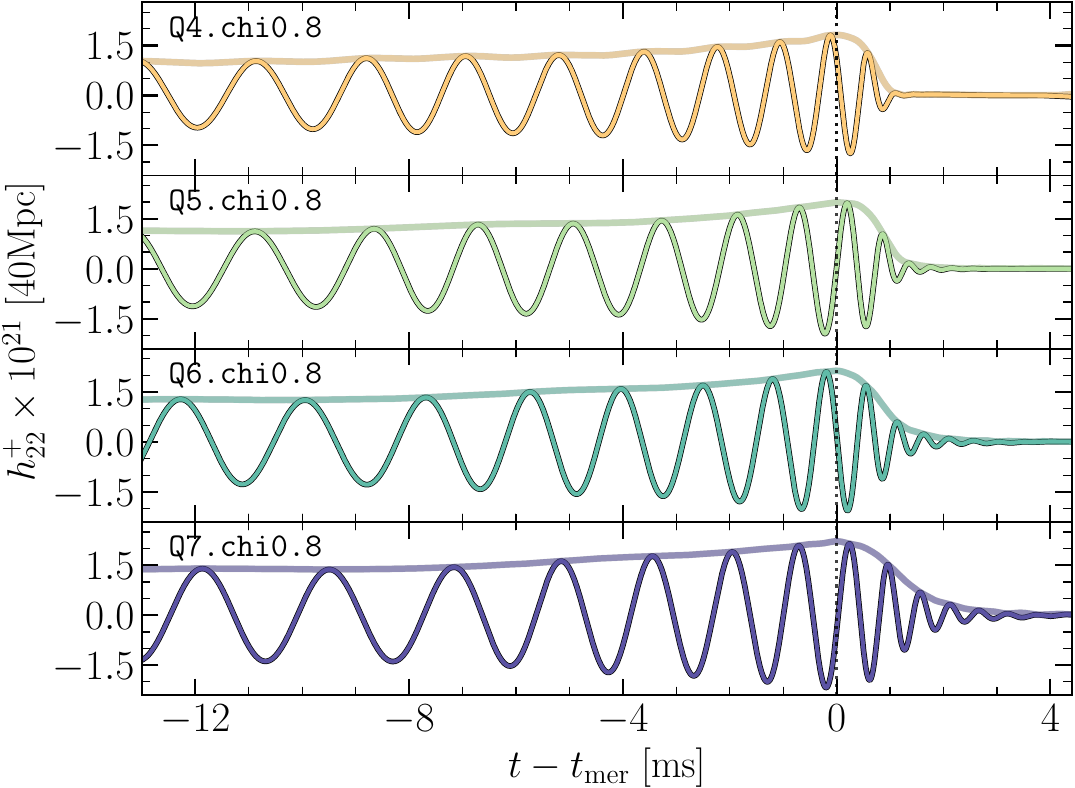}
  \caption{\textit{Left:} The $+$ polarization of the dominant $\ell=2=m$
    mode of the GW strain $h_{+}$, together with the total amplitude of
    that mode $\vert h \vert$, for BHNS systems with constant mass ratio
    ($Q=4$) and increasing BH spin from top to bottom ($\chi_{_{\rm BH}}
    = [0.0, 0.4, 0.6, 0.8]$). Dotted vertical lines serve as an
    additional indication for $t=t_{\rm mer}$ time. \textit{Right:} the
    same as on the left, but for binaries with fixed BH spin ($\chi_{_{\rm
    BH}}=0.8$) and increasing mass asymmetry from top to bottom ($Q = [4,
    5, 6, 7]$).}
  \label{fig:GW_q4567_chiBHs}
\end{figure*}

A simple argument can shed some light on this behaviour and reveal
potential caveats. We recall that the dimensionless spin is given by
$\chi_{_{\rm BH}}=S_{\rm BH}/M_{_{\rm BH}}^{2}$ and its differential
change can be expressed as $\delta\chi_{_{\rm BH}}=(1/M_{_{\rm
    BH}}^{2})\delta S_{\rm BH} - (2S_{\rm BH}/M_{_{\rm BH}}^{3}) \delta
M_{_{\rm BH}}$. Hence, the changes of the BH spin and mass are related to
the changes due to the accretion of angular momentum and mass via the
infalling matter. In the case of matter infalling with positive angular
momentum, as for all of our BHNS binaries, accretion will lead to an
increase of $\chi_{_{\rm BH}}$ via angular-momentum transfer and to a
decrease via mass transfer ($\delta S_{\rm BH} > 0$ and $\delta M_{\rm
  BH} > 0$). In binaries with larger $\chi_{_{\rm BH}}$, the angular
momentum transfer from the infalling matter is greater than in
irrotational binaries, causing larger gain in $S_{_{\rm BH}}$. On the
other hand, since the ISCO moves inward for higher prograde spins, matter
can occupy closer orbits without falling into the BH, hindering
accretion-induced spin-up. At the same time, if the disruption is more
prominent -- as is the case for BHs with increasing spins in binaries
with the same mass ratio -- the changes due to $\delta M_{_{\rm BH}}$
will be increasingly smaller [\cf Tab.~\ref{tab:sim_diag}]. The net result
between these competing effects depends on $Q$ and $\chi_{_{\rm BH}}$,
although the mass ratio effectively plays a secondary role in the
high-spin binaries considered here. All things considered, in all of the
BHNS binaries simulated in our work, the angular-momentum transfer always
dominates and the BH is always spun-up; however, it is perfectly possible
to spin-down the BH in binaries that have large enough $\chi_{_{\rm BH}}$
and small enough $Q$ (see Ref.~\cite{Lovelace2013} for an example). Also
in the case of final BH spin, a comparison with the analytic fitting
model suggested in Ref.~\cite{Zappa2019} reveals relative differences
that are smaller than $1\%$ in all the cases examined. Similar percent or
sub-percent differences are found when comparing our numerical values
with those presented in Ref.~\cite{Hayashi2023} for corresponding BHNS
binaries, thus once again confirming the high level of consistency
between the two numerical studies, but also the overall level of
precision with which the final BH remnant can be m0odeled in
numerical-relativity simulations.

\section{Results: gravitational-wave signal}
\label{sec:results_GW}

In this section we provide a detailed analysis of the GW signal extracted
from our simulations. As anticipated in Sec.~\ref{subsec:sim_diag_obs},
we extract Weyl scalar $\psi_4$ at a sphere of radius $r_{\rm
det}=600M_{\odot}\approx 886\, {\rm km}$ and from it obtain the GW strain
as defined in \eqref{eq:psi4}. In Fig.~\ref{fig:GW_q4567_chiBHs}, we show
the $+$ polarization of the GW strain $h_{+}$ along with its envelope,
the strain amplitude $\vert h \vert$, for all of the configurations
examined here with each signal shifted in time by $t_{\rm mer}$.

\subsection{Gravitational waveforms and their properties}

We begin by focusing on the configurations with fixed $Q=4$, which are
collectively shown in the left panel of Fig.~\ref{fig:GW_q4567_chiBHs},
and where it is possible to appreciate from top to bottom how the
influence of increasingly large prograde BH spin, which moves the ISCO
closer to the BH and facilitates tidal disruption, manifests itself with
a sharp decline in the signal amplitude after the merger. Indeed, in the
case of the binary \texttt{Q4.chi0.0}, for which no tidal disruption
takes place and the NS is absorbed as a whole, the GW signal after the
merger is essentially that of a perturbed BH, featuring the typical QNMs
of the corresponding BH. On the other hand, as the spin increases we
note a gradual quenching of both the inspiral and of the post-merger
signal. For the former part of the signal, this reduction is due to the
decreasing compactness of the NS matter as it is tidally disrupted,
hence leading to a reduction of the quadrupole moment of the system. For
the latter part of the signal, on the other hand, the quenching is
produced by the copious amount of matter that accretes onto the BH in an
almost axisymmetric manner.

\begin{figure*}
  \centering
  \includegraphics[width=0.47\textwidth]{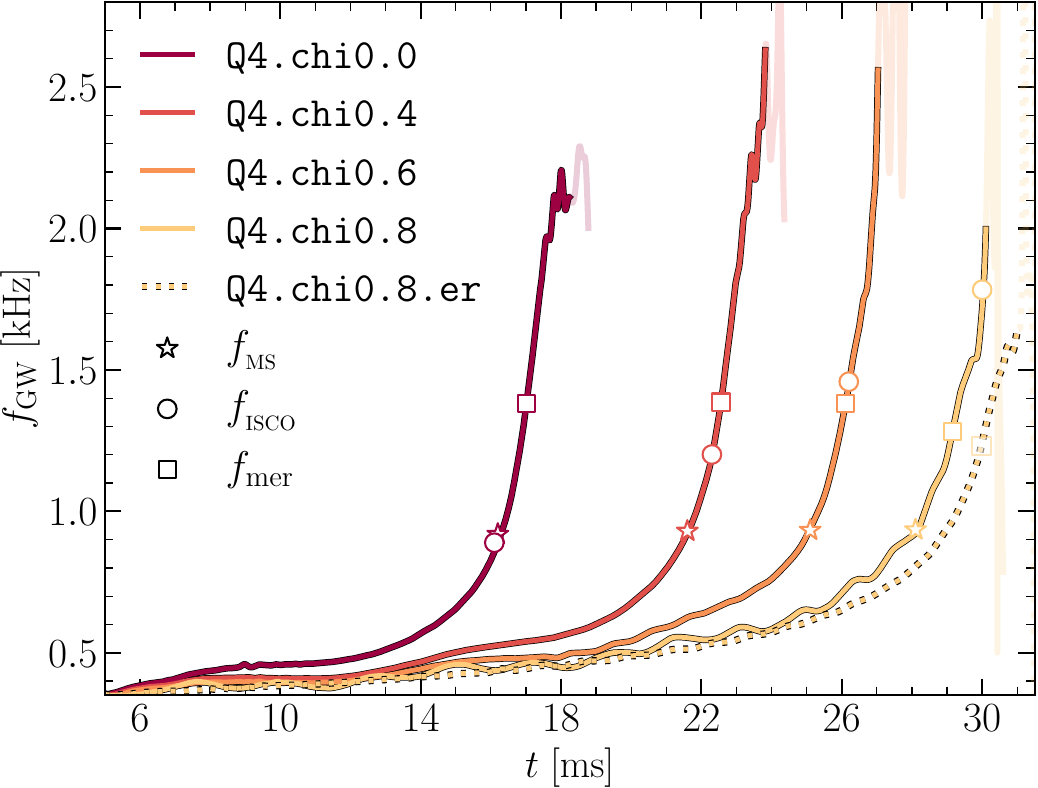}
  \hskip 0.5cm
  \includegraphics[width=0.47\textwidth]{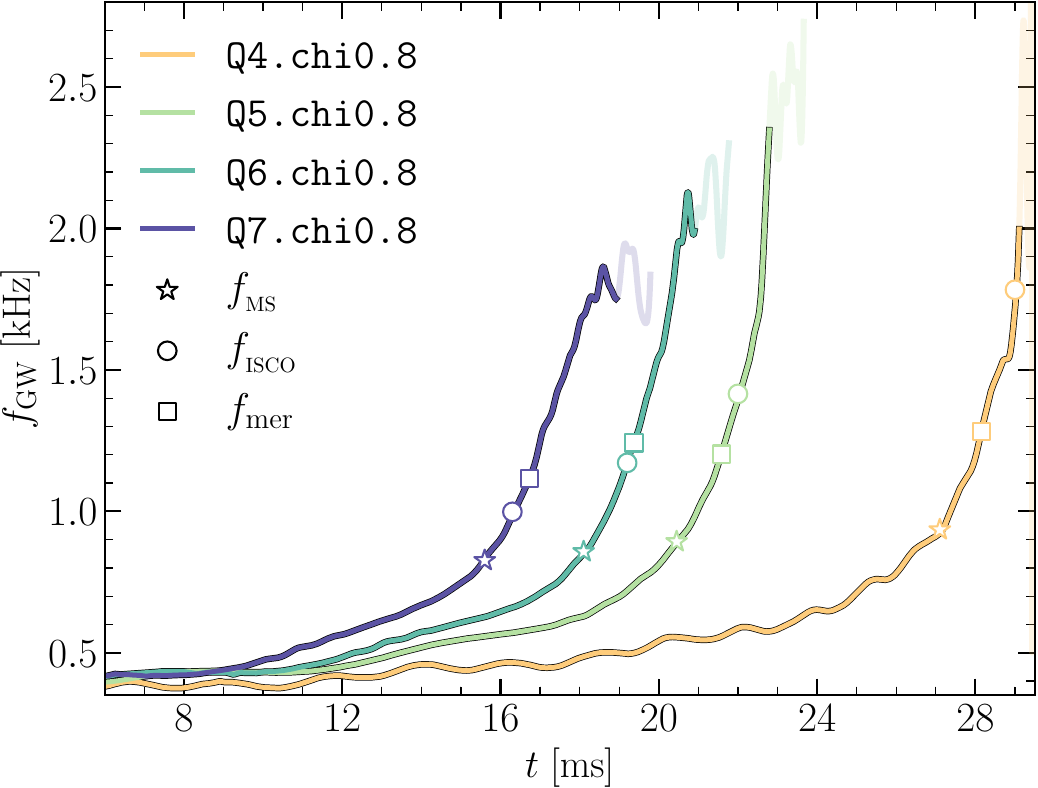}
  \caption{\textit{Left:} Evolution of the instantaneous GW frequency of
    the dominant $\ell=2=m$ mode for BHNS systems with constant mass
    ratio ($Q=4$) and increasing BH spin ($\chi_{_{\rm
    BH}} = [0.0, 0.4, 0.6, 0.8]$). Markers indicate the times when the GW
    reaches either the mass-shedding frequency $f_{_{\rm MS}}$ (stars) or
    the ISCO frequency $f_{_{\rm ISCO}}$ (empty circles). Also shown is
    the frequency at merger $f_{\rm mer}$ (empty squares).
    Semi-transparent line segments denote the transition of the signal
    from the inspiral to post-merger quasi-normal mode excitations. 
    Note that we also report the evolution of
    the eccentricity reduced binary \texttt{Q4.chi0.8.er} (dashed line).
    \textit{Right:} the same as on the left, but for binaries with
    fixed BH spin ($\chi_{_{\rm BH}}=0.8$) and increasing mass
    asymmetry ($Q = [4, 5, 6, 7]$).}
  \label{fig:bhns_fGW_q4567_chiBHs_wmark}
\end{figure*}

The right panel of Fig.~\ref{fig:GW_q4567_chiBHs}, on the other hand,
allows us to assess, in a simple manner and from top to bottom, the
impact that an increasing mass asymmetry $Q=[4, 5, 6, 7]$ has on the GW
signal when the BH spin is held constant to the large value of
$\chi_{_{\rm BH}}=0.8$. We recall that for a given BH spin, increasing
the BH mass will move the ISCO outwards (to greater separations),
and hence closer to the orbit at which the onset of mass shedding
occurs\footnote{As Paper I and the $f_{_{\rm MS}}$ column in
Tab.~\ref{tab:sim_diag} demonstrate, the impact of $M_{\rm Ch}$ on the
location of the mass-shedding orbit is smaller.}.
This is reflected in the properties of
the GW signal, which change significantly around merger time. Larger mass
asymmetries suppress tidal disruption and a clear, albeit distorted,
ringdown signal is present due to more structured, anisotropic matter
inflow. At the same time, because the mass of the BH increases with mass
asymmetry in the sequences that we have simulated, the damping time of
the ringdown signal increases visibly with $Q$. Thus, for the most
extreme mass ratio $Q=7$, the decay of the signal lasts approximately
$4\, {\rm ms}$ and fits several more oscillations than are present in 
the $Q=4$ case. The presence of delayed quenching of the post-merger
signal should be noted for the binaries \texttt{Q6.chi0.8} and
\texttt{Q7.chi0.8}, which is in contrast with what is seen for the
\texttt{Q4.chi0.8} binary. This behaviour can be attributed to the ISCO
being moved outwards for increasing BH mass. While the onset of mass
shedding still takes place outside of the ISCO (otherwise the merger
cannot be disruptive), the bound component of
matter which does not immediately fall into the BH necessarily finds
itself at a larger separation from the BH horizon and accordingly takes a
longer time to accumulate and quench the excitation of QNM. These
considerations and behaviours are in very good agreement with what
reported in similar studies, \eg in Refs.~\cite{Foucart2011,
Foucart2013a, Foucart2014, Kyutoku2015}.

\subsection{Spectral properties of the GW signal}
\label{subsec:gw_frequency}

Next, we analyze the properties of the GW signal in the frequency domain
so as to highlight the characteristic frequencies of the system and
ascertain their relation to the properties of the binary. We start by
reporting in Fig.~\ref{fig:bhns_fGW_q4567_chiBHs_wmark} the instantaneous
GW frequency $f_{_{\rm GW}}$ computed using the definition in
Eq.~\eqref{eq:fGW} and following the same convention for the ordering of
the binaries presented in Fig.~\ref{fig:GW_q4567_chiBHs} (\ie binaries
with constant mass ratio are shown on the left, while binaries with
constant spin are reported on the right). For clarity, each timeseries
ends roughly $1 \rm ms$ after the start of the ringdown and is shown with
a semi-transparent line segment to highlight the transition from the
exponential growth of the instantaneous frequency. In addition, for each
binary we use different markers to designate the frequency at merger
$f_{\rm mer}$ (colored boxes) and the frequencies computed from the QE
sequences that are related to the onset of mass shedding $f_{_{\rm MS}}$
(colored stars) and to the crossing of the innermost stable circular
orbit $f_{_{\rm ISCO}}$ (colored circles; see also
Appendix~\ref{sec:qe_sequences}). Also reported is the frequency
evolution for the eccentricity-reduced initial data
\texttt{Q4.chi0.8.er}; note that while the evolution in this case no
longer includes the modulations introduced by the eccentricity, the
relative difference between the merger frequencies of \texttt{Q4.chi0.8}
and \texttt{Q4.chi0.8.er} is $\ll 1\%$.

A number of basic features of the frequency evolution can be appreciated
in this way. More specifically, for binaries with constant mass ratio
(see left panel of Fig.~\ref{fig:bhns_fGW_q4567_chiBHs_wmark}):
\begin{itemize}
\item for all binaries considered $f_{_{\rm MS}} < f_{_{\rm ISCO}}$,
  hence indicating that a tidal disruption will occur and is observed.

\item the only exception to the rule above is for the \texttt{Q4.chi0.0}
  binary, for which $f_{_{\rm MS}} > f_{_{\rm ISCO}}$, and a plunge is
  indeed observed in this case.

\item the mass shedding frequencies $f_{_{\rm MS}}$ do not vary
  significantly with BH spin and are, approximately, constant.

\item the frequencies at the crossing of the ISCO $f_{_{\rm ISCO}}$ grow
  quadratically with the BH spin [see Eq.~(34) of paper I].

\item the merger frequencies $f_{\rm mer}$ do not show a systematic
  behaviour and are approximately constant.
\end{itemize}

\begin{figure}
  \centering
  \includegraphics[width=0.95\columnwidth]{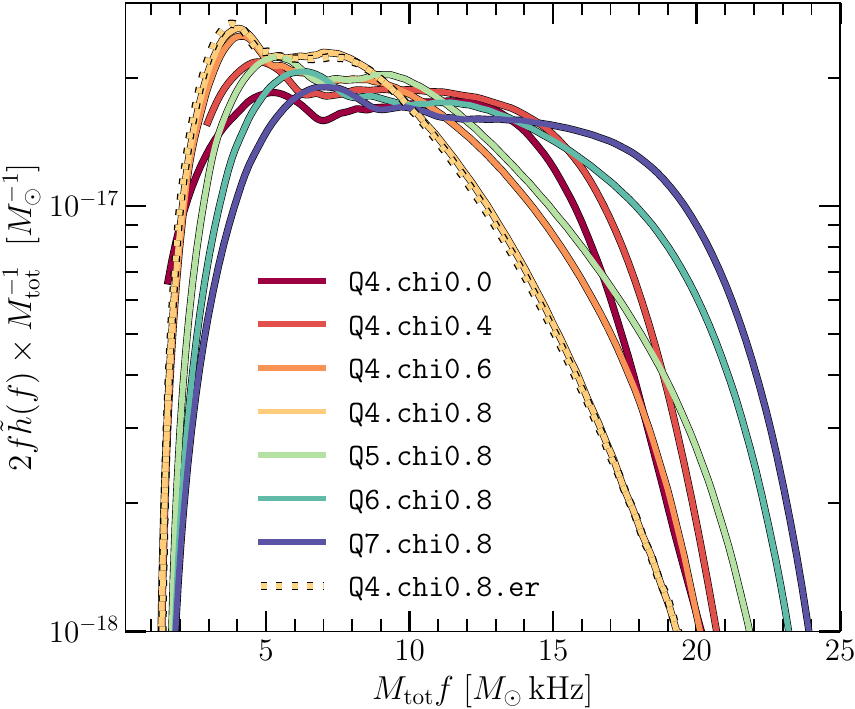}
  \hskip 0.5cm
  \caption{Power spectral densities $\tilde{h}(f)$ for all the binaries
    simulated in this work. The amplitude and the frequencies are
    rescaled by the total mass in order to facilitate comparison between
    different mass ratios and spins. Note that smaller mass asymmetries
    favour disruption, which contributes to the GW signal quenching at higher
    frequencies and reduces the power at lower frequencies via the
    reduced quadrupole moment of the system.}
  \label{fig:BHNS_GW_spectra}
\end{figure}

On the other hand, for binaries with constant BH spin (see right panel of
Fig.~\ref{fig:bhns_fGW_q4567_chiBHs_wmark}):
\begin{itemize}

\item the times in the evolution when the $f_{_{\rm MS}}$ and $f_{_{\rm
    ISCO}}$ frequencies are reached scale linearly with time.

\item neither the merger frequencies $f_{\rm mer}$, nor the times when
  these frequencies are reached, show a systematic behaviour and are
  approximately constant.
\end{itemize}

In an effort to make these considerations more quantitative, we define
the ``survival timescale'' $\tau_{\rm sur}$ as the difference between the
coordinate times when the instantaneous GW frequency reaches the mass
shedding and the marginally stable frequencies, $f_{_{\rm MS}}$ and
$f_{_{\rm ISCO}}$, \ie
\begin{equation}
  \tau_{\rm sur} := t(f_{_{\rm GW}}=f_{_{\rm ISCO}}) -
  t(f_{_{\rm GW}}=f_{_{\rm MS}})\,,
\end{equation}
then the survival time provides a measure of the interval between the
time when a fluid element is stripped from the NS and set into a stable
orbit and the time when it starts to accrete onto the
BH. Figure~\ref{fig:bhns_fGW_q4567_chiBHs_wmark} then essentially shows
that $\tau_{\rm sur}$ scales quadratically for binaries with constant
mass ratio and increasing BH spin, and linearly for binaries with
constant BH spin and increasing mass ratio. Even more interesting, while
$\tau_{\rm sur}$ is a quantity that can only be measured via
simulations, its value can be very well approximated using quantities
extracted from QE measurements. More specifically, it is possible to show
that the survival time can be very well approximated as
\begin{align}
  \label{eq:tau_sur}
  \tau_{\rm sur}  \simeq v_1 \, \tau^{^{\rm QE}}_{\rm sur}\,,
\end{align}
with the quasi-equilibrium survival time defined as
\begin{align}
  \label{eq:tau_sur_qe}
  \tau^{^{\rm QE}}_{\rm sur} :=
    g_{_{\rm ISCO}}(\chi_{_{\rm BH}}) \, \left(
      1/f_{_{\rm MS}} - 1/f_{_{\rm  ISCO}} \right) \,,
\end{align}
such that $\tau^{^{\rm QE}}_{\rm sur}$ is defined purely in terms of
binary characteristics ($Q, \chi_{_{\rm BH}}, M_{\rm tot}$) and
quantities derived from QE sequences. Here $g_{_{\rm ISCO}}(\chi_{_{\rm
    BH}})$ encapsulates the additional BH spin dependence of $\tau_{\rm
  sur}$ which accounts for the mild yet quadratic growth in $\tau_{\rm
  sur}$ for configurations with increasing BH spin [see Eq.~(34) in paper
  I and Tab.~\ref{tab:fitting_params_DD2} for the coefficients in the
  case of the DD2 EOS].
The fitting coefficient in Eq.~\eqref{eq:tau_sur} is given by $v_1 =
1.88$ and has a strong statistical regression of $R^2 = 0.988$. The
importance of expression Eq.~\eqref{eq:tau_sur} is that it allows one to
estimate the time between disruption and accretion avoiding expensive
full numerical-relativity simulations and relying instead uniquely on
quantities that can be computed via (comparatively) inexpensively QE
calculations.

Additionally, we show in Fig.~\ref{fig:BHNS_GW_spectra} the power
spectral density of the GW signal for all the binaries considered and
computed according to Eq.~\eqref{eq:PSD}. Note that the spectra and the
frequencies have been appropriately rescaled by the total mass of the
system to facilitate the comparison across different mass ratios and
spins~\cite{Lackey2013}. While this way of representing the data somewhat
obscures the actual frequencies involved, it allows for an easier
appreciation of the loss of power in the signal that is both mass ratio-
and spin dependent. For $Q=4$ configurations with varying spin, the
earlier quenching of the signal due to tidal disruption and suppression
of the black-hole quasi-normal modes is clearly visible. For example, in
units of $[M_{\odot} \, {\rm kHz}]$, the \texttt{Q4.chi0.0} binary has a
broad plateau from $5 \lesssim M_{\rm tot}\,f \lesssim 16$, with a steep
drop in power between $16 \lesssim M_{\rm tot}\,f \lesssim
20$. Conversely, the binary \texttt{Q4.chi0.8} lacks a plateau altogether
and instead has a more gradual decay in power from $8 \lesssim M_{\rm
  tot}\,f \lesssim 19$. Therefore, configurations where the matter inflow
is more structured and the outcome involves a less massive disk retain
more power in the high-frequency part of the spectrum corresponding to
the time around the merger.

In an effort to discover a quantitative method to classify the dynamics
of BHNS binaries, we leverage the qualitative classification explored in
Refs.~\cite{Kyutoku2011c, Pannarale2013a, Pannarale2015}, where three
main classes have been suggested. The first type of BHNS mergers is
obviously of the ``plunge'' type, where the signal is identical to a BBH
merger and mass-shedding occurs below the ISCO. This implies that there
is no trace of tidal disruption in the waveform. The second and the third
type further refine the classification of a binary that results in tidal
disruption. In the second, the merger is classified to have a
\textit{strong} disruption if there is a clear cut-off of the signal, \ie
a pronounced loss of amplitude related to tidal disruption at a distance
much larger than the ISCO. These binaries also result in massive
accretion disks. Lastly, the third type of the merger is the
\textit{weak} tidal disruption. In this case, which is characteristic for
high-mass asymmetries and large BH spins, the relative distance between
the mass-shedding orbit and the ISCO is small and a frequency cut-off is
less pronounced. At the same time, \textit{weak} disruptions will still
produce an accretion disk though with a smaller fraction of the NS mass
as compared to the binaries with a \textit{strong} disruption event.

Hence, according to the classification discussed above and with the help
of Fig.~\ref{fig:BHNS_GW_spectra} and the results described in previous
sections, we propose the following classification based on $\tau^{^{\rm
    QE}}_{\rm sur}$
\begin{align*}
  &\text{\rm ``plunge'':}     & 0  & \lesssim M_{\rm tot}^{-1} \, \tau^{^{\rm QE}}_{\rm sur}              \,, &&\\
  &\text{\rm ``weak disruption'':}   & 0  & \lesssim M_{\rm tot}^{-1} \, \tau^{^{\rm QE}}_{\rm sur} \lesssim 20 \,, &&\\
  &\text{\rm ``strong disruption'':} & 20 & \lesssim M_{\rm tot}^{-1} \, \tau^{^{\rm QE}}_{\rm sur}              \,. &&
\end{align*}
Using this quantitative classification scheme, the binary
\texttt{Q4.chi0.0} then belongs to the ``plunge'' class, while binaries
\texttt{Q4.chi0.4}, \texttt{Q4.chi0.6}, \texttt{Q6.chi0.8}, and
\texttt{Q7.chi0.8} should be classified as ``weak disruption''. Finally,
binaries \texttt{Q4.chi0.8} and \texttt{Q5.chi0.8} belong to the class of
``strong disruption''.  We have verified that this representation works
well for all the binaries we have considered and thus we propose it here
as an inexpensive and yet accurate manner of establishing the main
properties of the BHNS binary dynamics.

Before closing this section a few remarks are worth making. First, note
how the GW spectrum for the eccentricity-reduced binary overlaps
substantially with the non-reduced dataset, essentially because the
eccentricity in the latter is not substantial enough to be reflected in
the spectrum. Second, the \texttt{Q4.chi0.0} and \texttt{Q4.chi0.4}
binaries display a very similar high-frequency fall-off in power, albeit
more power is present for the \texttt{Q4.chi0.4} dataset for the
frequencies $M_{\rm tot}\,f \approx 20\,M_{\odot}\,{\rm kHz}$; this is
most likely due to the fact that the high-frequency part of the spectrum
is dominated by the ringdown, which is very similar in the two
binaries. Finally, binaries with fixed $\chi_{_{\rm BH}}=0.8$ and varying
mass ratio show a systematic increase of power at large frequencies as a
function of the increasing mass asymmetry. This is because the extent of
the tidal disruption decreases as a function of $Q$, which contributes to
the GW signal at higher frequencies.

\begin{figure*}
  \includegraphics[width=0.49\textwidth]{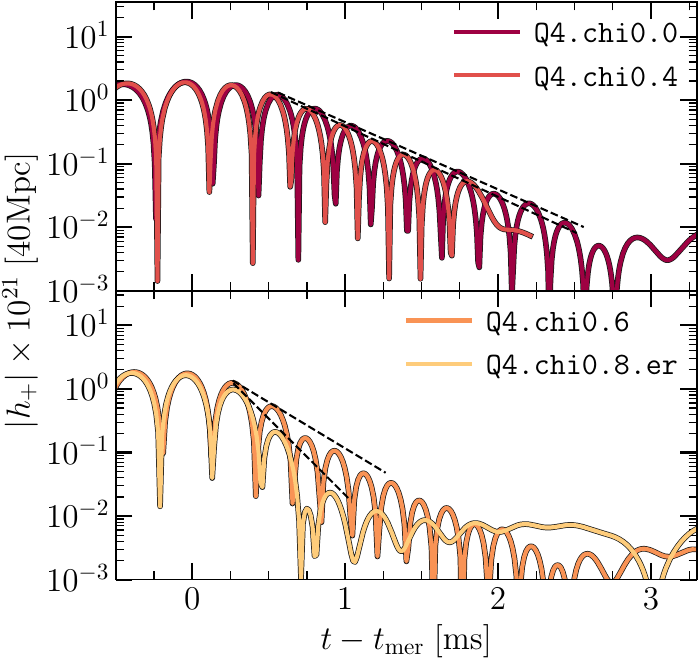}
  \hskip 0.25cm
  \includegraphics[width=0.49\textwidth]{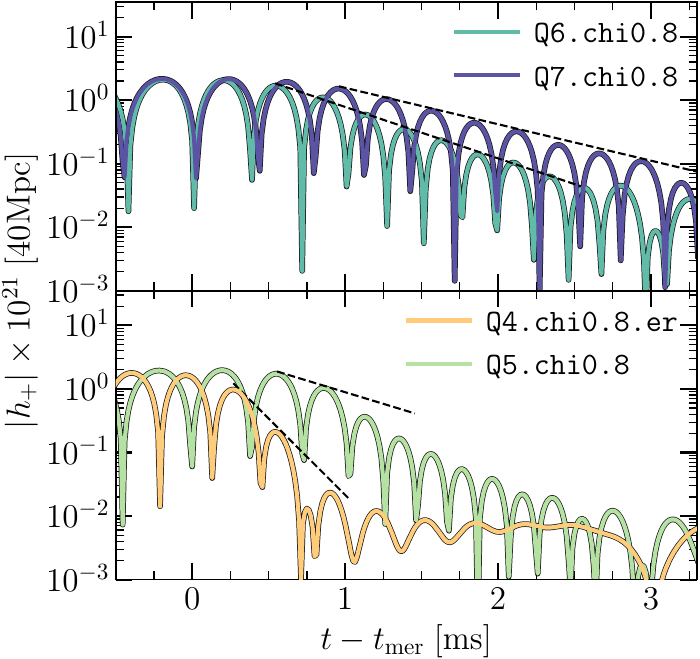}
  \caption{\textit{Left:} post-merger amplitudes of the $+$ polarization
    of the $\ell=2=m$ mode of the GW strain $h_{+}$. Reported are the GW
    signals from BHNS systems with constant mass ratio ($Q=4$) and
    increasing BH spin ($\chi_{_{\rm BH}} = [0.0, 0.4, 0.6, 0.8]$). The
    top panel shows binaries for which a genuine BH ringdown (see main
    text for a definition) is measured; the dashed lines mark the
    perturbative damping times of Kerr BHs having the same properties as
    the remnant BHs from the simulations, highlighting the very good
    match (see Tab.~\ref{tab:BH_qnm}). The bottom panel reports instead
    binaries for which a genuine BH ringdown cannot be found.
    \textit{Right:} the same as on the left but for binaries with fixed
    BH spin ($\chi_{_{\rm BH}}=0.8$) and increasing mass asymmetry ($Q =
    [4, 5, 6, 7]$). }
   \label{fig:bhns_amp_ringdown}
\end{figure*}
%

\subsection{Ringdown signal}
\label{sec:gw_ringdown}

Lastly, we turn our attention to the ringdown signal, \ie the portion of
the signal for $t>t_{\rm mer}$, which has a long history in the
literature of BH perturbations (see, \eg Refs.~\cite{Kokkotas99a,
  Nagar:2005ea, Berti:2009kk} for some reviews) and has been studied,
albeit to a much lesser extent, also for BHNS
mergers~\cite{ShibataTaniguchi2008, Lackey2013, Pannarale2015,
  Gonzalez2022a}. Following the notation of Ref.~\cite{Berti:2009kk}, we
briefly recall that a perturbed BH (as the one produced by the merger of
a BBH or BHNS binary) leaves behind a remnant BH in an excited state,
which the black hole ''sheds'' through the emission of GWs. In
Schwarzschild or Kerr spacetimes, which represent the asymptotic final
states of a BBH merger~\cite{Hofmann2016}, the excitations can be
decomposed as a sum of QNMs indexed by three integers $(\ell, m,n)$, each
with its unique frequency $f_{\ell m n }$ and damping time-scale
$\tau_{\ell m n}$, where $\ell,m$ are related to the decomposition in
terms of spin-weighted spherical harmonics $^{s}Y_{\ell m}$ with weight
$s=-2$, and $n$ denotes the \textit{overtones}, with $n=0$ the
fundamental mode.

Focusing solely on the $+$ polarization of the fundamental frequency
$n=0$ of the dominant mode $(\ell, m)=(2,2)$, its contribution to the
signal is given by
\begin{equation}
  (h_{+})_{\rm rd} =\frac{M_{_{\rm BH}}}{r} \Re
  \left[A^{+}_{220}e^{i(\omega_{220}t + \phi_{220})}e^{-t/\tau_{220}}\right]\,,
  \label{eq:QNM_Kerr}
\end{equation}
where $A^{+}_{220}$ is the amplitude of the mode, $\phi_{220}$ is the
phase shift, $r$ the distance from the BH and $\tau_{220}$ the damping
timescale. For any given mode, the numerical values of the complex
eigenfrequency $\omega_{_{\rm QNM}} := \omega_{_{R}} + i \omega_{_{I}}$,
of the ``quality factor'' $\mathcal{Q} := {\omega_{_{R}}}/{2
  \omega_{_{I}}}$, and of the damping time $\tau_{d}:=1/\omega_{_{I}}$
can be calculated from fits of numerical perturbative calculations. In
the case of the $(220)$ mode, these values (truncated at the second
significant figure) are given by~\cite{Berti06b}
\begin{align}
  \label{eq:QNM_freq}
  M_{_{\rm BH}}\omega_{220} &= 1.53 - 1.16\, \left(1-\chi_{_{\rm BH}}\right)^{0.13}\,, &\\
  \label{eq:QNM_damp}
  \mathcal{Q}_{220}     &= 0.70 + 1.42\, \left(1-\chi_{_{\rm BH}}\right)^{-0.50}\,. &
\end{align}
and provide a very good approximation of the ringdown observed, for
instance, in BBH mergers.

Crucial for the validity of the expressions \eqref{eq:QNM_freq} and
\eqref{eq:QNM_damp} above, that are obtained using linear-perturbation
theory, is that the mass and spin of the BH do not change during the
ringdown. However, for BHNS mergers, the presence of a companion NS of
finite tidal deformability will impact the ringdown signal in a number of
ways if a tidal disruption occurs. Indeed, the properties of the remnant
BH will change either via the inflow of the tidally disrupted NS --
which, as discussed above, changes the BH mass and spin -- or via the
formation of an accretion disk -- so that the spacetime is no longer well
approximated by a Schwarzschild or Kerr solution.

Under these conditions, it is essential to recognize which binaries
produce a clear ringdown signal and which ones do not. Indeed, assuming
that there is a ringdown signal as predicted by the perturbative result
in Eq.~\eqref{eq:QNM_Kerr} when none is present, or trying to model the
signal in terms of the perturbative prediction \eqref{eq:QNM_Kerr} when
the signal differs considerably from it, can lead to significant errors
in the inference of the properties of the merger remnant, not only in
terms of the BH mass and spin, but also in terms of the remnant disk. In
view of these considerations, we make the \textit{working} definition
that a genuine BH ringdown is present in the post-merger signal if and
only if the latter shows right after the amplitude maximum a clear
exponential fall-off and the ratio between the measured quality factor
and the one derived from BH perturbation theory is $\mathcal{Q} /
\mathcal{Q}_{\rm Kerr} \gtrsim 0.8$. Stated differently, we consider as
genuine BH-ringdown a GW signal that, right after the merger, shows
sufficiently many oscillations having the same period and whose amplitude
decreases exponentially, taking the Kerr ringdown as a reference. We note
that this definition would not consider as a genuine BH ringdown a signal
that follows the perturbative behaviour in Eq.~\eqref{eq:QNM_Kerr}, but
that starts much later than the merger time.

Figure~\ref{fig:bhns_amp_ringdown} collects the post-merger signal of the
$h_{+}$ strain for all of the BHNS binaries considered here. As in the
preceding figures of this type (\ie ..~\ref{fig:GW_q4567_chiBHs} and
\ref{fig:bhns_fGW_q4567_chiBHs_wmark}), the left panel shows the impact
of spin at constant mass ratio ($Q=4$), while the right one the influence
of the mass ratio for fixed BH spin ($\chi_{_{\rm BH}}=0.8$). For
clarity, we report the ringdown of the eccentricity-reduced binary
\texttt{Q4.chi0.8.er} rather than that of \texttt{Q4.chi0.8} so as to
highlight the contribution of the eccentricity on the ringdown (see
Appendix~\ref{sec:ecc_ringdown} for a comparison between the two).

Note that each panel in Fig.~\ref{fig:bhns_amp_ringdown} reports two
sub-panels, where we have collected in the top sub-panels the BHNS
binaries that produce a clear ringdown as defined above, while we show in
the bottom sub-panels the binaries whose post-merger signal does not
feature the properties we expect in a genuine BH ringdown. Hence, the
BHNS binaries \texttt{Q4.chi0.0}, \texttt{Q4.chi0.4}, \texttt{Q6.chi0.8},
and \texttt{Q7.chi0.8} exhibit a clear ringdown, while the other binaries
have post-merger signals that are either very different from a BH
ringdown, \ie \texttt{Q4.chi0.8}, or simply fail to satisfy our working
definition of BH ringdown, \ie binaries \texttt{Q4.chi0.6} and
\texttt{Q5.chi0.8}. Interestingly, such binaries do show a signal that is
\textit{almost} periodic and that is decaying \textit{almost}
exponentially, although not sufficiently periodic or without a single
exponential fall-off. Binaries of this type may be used to build
effective models that are constructed around the perturbative
expression~\eqref{eq:QNM_Kerr}, but are distinct from the aforementioned
ringdown picture (see~\cite{Pannarale2013a, Pannarale2015} for some first
phenomenological approach in this sense). Such models would need to be
carefully tested to guarantee that their use leads to a faithful
reconstruction of the merger-remnant properties (see, \eg the post-merger
signal reconstruction for binaries with a strong disruption in the upper
panels of Fig.~4 in Ref.~\cite{Gonzalez2022a}).

We can use these results for the ringdown to close the loop around the
considerations made regarding the degree of disruption in
Sec.~\ref{subsec:gw_frequency}. Interestingly, in fact, the binaries that
do match our working definition of genuine BH ringdown are also those
that fall under the class of ``no disruption'' or ``weak disruption'';
conversely, the binaries that fail our definition, fall under the class
of ``strong disruption''. This internal consistency can be used to make a
prediction and, in particular, to conjecture that BHNS binaries with
$M^{-1}_{\rm tot}\,\tau^{^{\rm QE}}_{\rm sur} \lesssim 20$ lead to a
post-merger signal with a clearly identifiable BH ringdown signal.

\begin{table*}[t]
  \begin{ruledtabular}
    \begin{tabular}{l|cccc|cc|ccc}
	    binary & $M\omega_{_{R}}$ & $M\omega_{_{I}}$ &
	    $f_{\rm QNM}$ & $\tau_{d}$ &
	    $M\omega_{_{R}}^{\rm Kerr}$ & $M\omega_{_{I}}^{\rm
	    Kerr}$ & $\mathcal{Q}$ & $\mathcal{Q}_{\rm Kerr}$ & $\mathcal{Q}/\mathcal{Q}_{\rm Kerr}$ \\
	     & & & $[$kHz$]$ & $[{\rm
	    ms}]$ & & & & & \\
      \hline
	    \texttt{Q4.chi0.0}              & $0.467$ & $0.088$ & $2.169$ & $0.389$ & $0.460$ & $0.087$ & $2.653$  & $2.643$  & $1.004$\\
	    \texttt{Q4.chi0.4}   	    & $0.491$ & $0.085$ & $2.317$ & $0.405$ & $0.526$ & $0.082$ & $2.888$ & $3.207$ & $0.901$\\
	    \texttt{Q4.chi0.6}$^{\star}$    & $0.558$ & $0.106$ & $2.720$ & $0.308$  & $0.569$ & $0.078$ & $2.632$ & $3.647$ & $0.722$ \\
	    \texttt{Q4.chi0.8}$^{\star}$    & $0.583$ & $0.124$ & $2.857$ & $0.261$  & $0.636$ & $0.069$ & $2.351$ & $4.609$ & $0.510$ \\
	    \texttt{Q5.chi0.8}$^{\star}$    & $0.426$ & $0.095$ & $1.724$ & $0.473$  & $0.627$ & $0.070$ & $2.235$ & $4.478$ & $0.499$\\
	    \texttt{Q6.chi0.8}              & $0.625$ & $0.083$ & $2.147$ & $0.557$  & $0.626$ & $0.070$ & $3.765$ & $4.471$ & $0.842$ \\
	    \texttt{Q7.chi0.8}              & $0.615$ & $0.070$ & $1.856$ & $0.764$  & $0.626$ & $0.070$ & $4.393$ & $4.471$ & $0.982$\\
      \hline
	    \texttt{Q4.chi0.8.er}$^{\star}$ & $0.583$ & $0.178$ & $2.857$ & $0.182$  & $0.636$ & $0.069$ & $1.638$ & $4.609$ & $0.355$
    \end{tabular}
  \end{ruledtabular}
  \caption{Estimates of $M\omega_{_{R}}$, $M\omega_{_{I}}$, $f_{\rm QNM}$
    and, $\tau_{d}$ in the post-merger signal for all of the simulations
    in this work; binaries without a genuine ringdown are marked with an
    asterisk and we set $M=M_{\rm Ch}$ for compactness. The last two
    columns report the corresponding eigenfrequencies $\omega_{_{R}}^{\rm
      Kerr}$ and $\omega_{_{I}}^{\rm Kerr}$ relative to a BH with mass
    $M_{\rm Ch}^{\rm rem}$ and spin $\chi_{_{\rm BH}}^{\rm mer}$ [see
      Eqs.~\eqref{eq:QNM_freq} and~\eqref{eq:QNM_damp}]. Additionally, we
    report the measured quality factor $\mathcal{Q}$, the one obtained
    from perturbative theory $\mathcal{Q}_{\rm Kerr}$ for a BH with
    $\chi_{_{\rm BH}}^{\rm mer}$ ($\mathcal{Q}_{\rm Kerr}$ is a function
    of the spin only), as well as their ratio that measures how close our
    measured ringdown is to the expected one for an isolated Kerr
    BH. Dashed lines in Fig.~\ref{fig:bhns_amp_ringdown} clearly indicate
    that the damping strength increases after the two initial peaks.}
\label{tab:BH_qnm}
\end{table*}

Overall, in contrast to strongly disruptive mergers for which the signal
ends abruptly, the cases with a clear ringdown demonstrate agreement with
perturbation theory not only in terms of the damping timescale, but also
the frequency of QNM. To that end, we find that in spite of almost
identical remnant BH masses for the \texttt{Q4.chi0.0} and
\texttt{Q4.chi0.4} binaries (the relative difference $\lesssim 0.25\%$),
the real part of the ringdown eigenfrequency differs. Indeed, the higher
spin of the BH remnant for the latter configuration is reflected in a
higher frequency of the ringdown, as expected from perturbation theory
[\cf Eq.~\eqref{eq:QNM_freq}]. This excellent consistency further
justifies the rigorous criteria for a ringdown that we have
employed. Furthermore, while \texttt{Q4.chi0.6} and \texttt{Q4.chi0.8} do
not satisfy our definition of a ringdown, since some excitation of QNM is
nevertheless present, our estimates of $\omega_{_{R}}$ indicate that it
increases further, consistent with a higher remnant BH spin. Measurements
of $\omega_{_{R}}$ and $\omega_{_{I}}$ obtained by fitting
Eq.~\eqref{eq:QNM_Kerr} to a suitably regular part of the post-merger
signal are reported in Tab.~\ref{tab:BH_qnm} for all of the binaries here
and compared with the perturbative expectation revealing relative
differences between $1\%$ and $6\%$ for ringdown cases.
The post-merger signals without a genuine ringdown are marked
with a star and the corresponding values of $\tau_{d}$ are to be
understood as upper bounds on the damping timescale, since the signal is
damped stronger than exponentially. Furthermore, the inspection of the
ratio of the quality factors $\mathcal{Q}/\mathcal{Q}_{\rm Kerr}$
measures the deviation of the ringdown from Kerr (which we require to be
at least $80\%$ for a genuine ringdown) and that $\omega_{_{I}} >
\omega_{_{I}}^{\rm Kerr}$ for strong-disruption cases. Importantly, we
stress that Fig.~\ref{fig:bhns_amp_ringdown} presents $\vert h_{+}
\vert$, \ie the positive and negative part of one polarization.
Therefore, the frequencies $f_{\rm QNM}$ as listed in
Tab.~\ref{tab:BH_qnm} correspond to every \textit{second} peak of that
quantity.

As a final remark on this section, we note that the reason why the
binaries \texttt{Q4.chi0.6} and \texttt{Q4.chi0.8} fail spectacularly to
exhibit a BH ringdown signal is because in these cases, the BH undergoes
significant changes in mass and spin over rather large timescales [see
  ..~\ref{fig:BHNS_sim_BH_Chmass} and \ref{fig:BHNS_BHspin} and the
  corresponding discussion]. Under these conditions, which are very
different from those assumed when studying BH-perturbation theory, the
ringdown signal is quenched by the significant inflow of matter and
angular momentum -- that effectively acts as an incoherent excitation of
the BH -- and the BH response is more similar to that of an over-damped
oscillator rather than that of a free damped oscillator.

\section{Conclusions}
\label{sec:summary}

Using fully general-relativistic GRMHD simulations of BHNS mergers, we
have carried out a systematic investigation of the region of the space of
parameters leading either to a plunge or to a tidal disruption, with the
latter being of particular interest for multi-messenger astronomy. In
particular, in this second paper in a series, we have explored the binary
dynamics when varying the mass ratio and the BH spin while keeping fixed
a realistic and temperature-dependent EOS and a reference NS mass.

One of the highlights of the analysis carried out in this work is the
characterization of tidal disruption by means of a suitable contraction
of the Riemann tensor, evaluated in a frame comoving with the fluid; to
our knowledge, this constitutes the first instance when its computation
is performed in a dynamical simulation. We have also demonstrated that
this derived scalar quantity displays exponential growth at the onset of
tidal disruption and hence presents a complementary picture that can be
used to understand the development of instability that leads to unstable
mass transfer onto the black hole; the growth rate can be extracted to
unambiguously infer the transition from mass shedding to complete loss of
a coherent NS structure.

Particularly worthy of note in our set of BHNS binaries are those
featuring a mass asymmetry $Q = 6, 7$ and a BH of maximal spin admissible
by the initial data solver, $\chi_{_{\rm BH}}=0.8$, which simultaneously
constitute the most challenging configurations here, are rarely covered
by numerical investigations, and are on-par with the most extreme ones
reported in the literature.

Special attention has been paid to the dynamics of the matter undergoing
disruption and we have discussed in detail the spatial extent of the
early post-merger disks, their density and temperature distributions, as
well as the properties of the remnant BHs. Once again, of particular
interest is the $Q=7$ run, which corresponds to the currently conjectured
peak in the mass ratio distribution among BHNS systems. Despite this
appreciable mass ratio, the large prograde BH spin still leads to a
substantial amount of matter both in the bound and unbound components.
We explicitly show that this is reflected in the timescale
for the formation of the accretion disk in the fallback
accretion rate. We have also found that existing analytic
fitting models for the BH remnant and total remnant
rest-mass~\cite{Foucart2018b} show rather large disagreements with the
results of fully-relativistic simulations in the regime of large mass
asymmetry and high BH spin, whereas the simulations across different
groups in said regime display a surprising but welcome consistency. This
finding highlights the need for a broader coverage of the parameter space
of high mass asymmetry BHNS binaries with significant BH spins to be
employed for the construction of more accurate analytical models.

Our work has also investigated the influence of mass ratio and BH spin on
the GW signal and its frequency spectrum. From this we naturally recover
the known influence of tidal disruption on suppressing the signal's
amplitude and quenching the excitation of QNMs. In particular, we have
demonstrated that a number of post-merger signals in our work display a
behaviour that is consistent with the one expected of a BH ringdown in
the context of perturbation theory. This is the case for those binaries
that either lead to a plunge or where the tidal disruption is
``weak''. For these cases, we have also measured the QNM frequencies and
damping timescales relative to the asymptotic BH properties and found
differences between $1\%$ and $6\%$. On the other hand, we have also
shown that binaries exhibiting ``strong'' disruption episodes lead to
considerably different post-merger signal that has little resemblance
to a BH ringdown. This result highlights the importance of proper
modelling of the post-merger GW signal in those BHNS binaries with
moderate mass asymmetries ($Q=4-5$) and involving rapidly rotating BHs
($\chi_{_{\rm BH}} \gtrsim 0.8$).

An entirely novel part of our analysis involves a consistency check
between the predictions of QE sequences presented in paper I and fully
dynamical simulations. In addition to validating the ability for QE
estimates to predict whether a BHNS binary will lead to a plunge or a
tidal disruption, and for the first time with the inclusion of terms
modelling BH spin dependence,
we have confirmed that the relation between the QE
frequencies at mass shedding and at the effective ISCO, $f_{_{\rm MS}}$
and $f_{_{\rm ISCO}}$, correspond accurately to the relativistic outcome
of the merger. The difference between these frequencies decreases as the
NS is disrupted closer to the ISCO radius, thus producing a smaller
remnant disk and with the post-merger GW signal being closer to a BH
ringdown. In the limit of our plunge configuration \texttt{Q4.chi0.0},
where $f_{_{\rm MS}} \gtrsim f_{_{\rm ISCO}}$, we find a negligible
amount of material in the post-merger remnant and a GW signal very close
to that from a BBH merger, as expected from paper I. Finally, we find a
novel relation between the dynamical time $\tau_{\rm sur}$ which
is strongly correlated to $f_{_{\rm ISCO}}$ and $f_{_{\rm MS}}$. From
this relation and the measurements of $\tau_{\rm sur}$, we provide for
the first time an NR-validated criterion based solely on sequences of quasi-equilibrium
initial data ($\tau^{^{\rm QE}}_{\rm sur}$) to indicate whether a BHNS
merger will result in a plunge, in a weak or in strong tidal disruption.

As a concluding remark, we note a number of different directions that
could be explored in future works. For a significantly extended
timescale, the inclusion of neutrino physics and stronger magnetic fields
is required to dictate the dynamics of the post-merger accretion disk and
the asymptotic state of the tidal tail accurately. In particular,
accurate modelling of weak interactions is crucial in determining the
composition and temperature of dynamical ejecta, which could then be used
as input for nucleosynthetic evolution by means of nuclear reaction
networks. The corresponding results could then be used to estimate the
magnitude and longevity of the fission-driven kilonova emission. It would
be of considerable value if nucleosynthesis and kilonova emission could
be correlated to QE predictions as sequences of QE initial data are on
average a factor ten less expensive to compute and at least four times
faster to compute when only considering the average time and resources to
compute the inspiral and merger for the configurations considered here.
Finally, while simulations with the mass asymmetry $Q=7$ are some of the
most extreme ones investigated in BHNS simulations, they do not yet reach
the tail of the conceivable distribution among the BHNS systems expected
in nature \cite{Broekgaarden2021}. In particular, the effort of providing
\textit{fully relativistic} predictions for BHNS systems with
$Q>8.3$~\cite{Kyutoku2021a} and featuring strong or weak disruption,
accompanied by a comparison with the existing remnant
mass~\cite{Foucart2018b} and ejecta mass models~\cite{Kyutoku2015} has
not yet been explored.

\section*{Data Availability}

The general diagnostics, scalar timeseries and initial data for the
dynamical simulations used in preparing this article can be shared upon
a reasonable request to the corresponding author.

\begin{acknowledgments}
KT gratefully acknowledges the extensive discussions with R. Duqu\'e in
the lead up to this project. Support in funding comes from the State of
Hesse within the Research Cluster ELEMENTS (Project ID 500/10.006), from
the ERC Advanced Grant ``JETSET: Launching, propagation and emission of
relativistic jets from binary mergers and across mass scales'' (Grant
No. 884631). The computation of quasi-equilibrium sequences and BHNS
simulations were performed on HPE Apollo HAWK at the High Performance
Computing Center Stuttgart (HLRS) under the grants BNSMIC and BBHDISKS.
ST gratefully acknowledges support from NASA award ATP-80NSSC22K1898.
\end{acknowledgments}

\bibliographystyle{apsrev4-2}
%

\appendix

\section{Predictions from quasi-equilibrium sequences}
\label{sec:qe_sequences}

To aid the comparison with the QE analysis and predictions made in paper
I~\cite{Topolski2024b}, we here report the most essential results
presented there, as well as draw several parallels between the
predictions of QE sequences and the dynamical results. To the best of our
knowledge, this constitutes the first effort to describe the dynamics of
tidal disruption both in terms of dynamical simulations and in terms of
QE analyses.

We start by recalling the important definitions and diagnostics of paper
I to facilitate the comparison with the results of GRMHD simulations. To
ensure a meaningful comparison, we have generated initial-data sequences
with the DD2 EOS that we use in this paper assuming a fixed NS mass
$M_{_{\rm NS }}=1.4 \, M_{\odot}$. While the results of paper I suggest a
degree of quasi-universality in many of the QE properties, computing the
QE sequences for the DD2 EOS used in the dynamical evolutions ensures
that we minimize any potential discrepancy that could affect our
comparison. To that end, we have computed a restricted set of QE sequences
that spans mass ratios $Q=[4, 5, 6]$ and BH spins $\chi_{_{\rm BH}}=[0.0, 0.4,
0.8]$, covering precisely the range of the parameter space that we wish
to explore using numerical-relativity simulations. We then employ
the techniques and procedures outlined in paper I to obtain
characteristic orbital angular velocities at the innermost stable
circular orbit $\Omega_{_{\rm ISCO}}$ and at the onset of mass shedding
$\Omega_{\rm MS}$, represented as functions of the stellar compactness
$\mathcal{C}$, inverse mass ratio $Q$, and black hole spin $\chi_{_{\rm
BH}}$. We recall from paper I that the full $(\mathcal{C}, Q, \chi_{_{\rm
BH}})$ dependence of these frequencies can be expressed analytically as
\begin{align}
  \label{eq:mtotomega_fits_ms}
  M_{\rm tot}\Omega_{_{\rm MS}}   &:=  c_{1}
  \mathbin{\color{black!90!black}
    \mathcal{C}^{3/2} \left(1+Q\right) \left(1 + 1/Q \right)^{1/2} } \\
  &\phantom{=} \times
  (1+c_{2} \, \mathcal{C})(1+c_{3} \, Q) \,,\nonumber \\
  &\phantom{=} \times (1 + e_{_{1,\rm MS}} \chi_{_{\rm BH}} +
  e_{_{2,\rm MS}}\chi_{_{\rm BH}}^{2}) \nonumber \\
  &\phantom{=} \times (1 + e_{_{3,\rm
      MS}} \chi_{_{\rm BH}}Q) \nonumber \\
  \label{eq:mtotomega_fits_isco}
  M_{\rm tot} \Omega_{_{\rm ISCO }}&:= \mathbin{\color{black!90!black}
    6^{-3/2} }\left(1 + d_{1}\, Q^{-d_{2}}  \right)
  \left( 1 + d_{1} \, \mathcal{C}^{d_{3}}/Q \right) \\
  &\phantom{=} \times (1 + e_{_{1,\rm ISCO}} \, \chi_{_{\rm BH}} +
  e_{_{2,\rm ISCO}} \,\chi_{_{\rm BH}}^{2}) \,, \nonumber
\end{align}
where the numerical values for the best-fit coefficients for the DD2 EOS
are presented in Tab.~\ref{tab:fitting_params_DD2}.

In Fig.~\ref{fig:Om_Dis_Cond_DD2} in the main text we present, as a
function of the BH spin and NS radius, the marginal disruption condition
$\Omega_{_{\rm MS}}=\Omega_{_{\rm ISCO}}$ in the form of a critical mass
ratio $Q_{\rm crit}$ that separates the regions of ``tidal disruption''
and ``plunge'' scenarios for a fixed neutron star masses $M_{_{\rm
NS}}=1.4\, M_{\odot}$. If the mass ratio of the system is smaller than
$Q_{\rm crit}$ (\ie the BH mass is smaller than $Q_{\rm crit }M_{_{\rm
NS}}$) for a given $R_{_{\rm NS}}$ and $\chi_{_{\rm BH}}$, then mass
shedding will begin before the NS encounters the ISCO.
Figure~\ref{fig:Om_Dis_Cond_DD2} should be contrasted with Fig.~13 in
paper I, where we have employed three different EOSs and where the same
$Q_{\rm crit}$ contours are moved to lower values of spin. This is
because the DD2 EOS employed here is stiffer and leads to smaller
compactnesses for the same radius, making the NS easier to disrupt.

The contours of critical mass ratio $Q_{\rm crit}$ reported in
Fig.~\ref{fig:Om_Dis_Cond_DD2} from the QE analysis clearly show that all
of the configurations aside from \texttt{Q4.chi0.0} (see
Sec.~\ref{subsec:param_space}) should lead to a tidal disruption; this
prediction is indeed confirmed by the dynamical simulations, where tidal
disruptions always take place, although with different strengths. The
only exception signalled by Fig.~\ref{fig:Om_Dis_Cond_DD2} is for the
\texttt{Q4.chi0.0} binary that is below the corresponding $Q_{\rm crit} =
4$ contour and should therefore not be disrupted, but experience a plunge
instead. Indeed, also in this case, the dynamical simulations reveal that
this is exactly what happens, thus confirming the robustness of the
predictions of the QE calculations and highlighting the synergies that
are possible when combining the QE analysis -- that, in principle, can be
repeated for any EOS with the open-source code \fuka~ -- with fully
dynamical simulations.

\begin{table}[t]
  \begin{tabular}{|l|c|c|c|}
	  \hline \hline
    EOS & DD2 &  & \\
	  \hline
	  $c_{i}$ & $\phantom{-}0.369$ & $-0.999$ & $-0.020$ \\
	  \hline
	  $d_{i}$ & $\phantom{-}0.793$ & $\phantom{-}0.676$ & $\phantom{-}0.484 $ \\
	  \hline
	  $e_{i, \rm MS}$ & $\phantom{-}0.057$ & $-0.020$ & $-0.005$ \\
	  \hline
	  $e_{i, \rm ISCO}$ & $\phantom{-}0.497$ & $\phantom{-}0.951$ & $-$\\
    \hline \hline
  \end{tabular}
  \caption{Fitting parameters for the $\Omega_{_{\rm MS}}$ and
    $\Omega_{_{\rm ISCO}}$ modelling functions in
    Eq.~\eqref{eq:mtotomega_fits_isco} and
    Eq.~\eqref{eq:mtotomega_fits_ms}, for the DD2 EOS.}
   \label{tab:fitting_params_DD2}
\end{table}

\begin{figure}[t!]
  \centering
  \includegraphics[width=0.95\columnwidth]{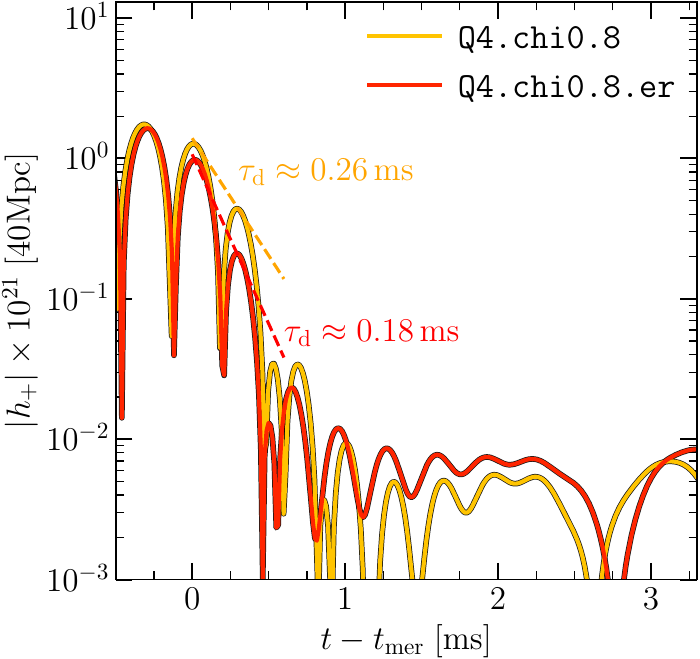}
  \caption{Post-merger signal for the \texttt{Q4.chi0.8} and
    \texttt{Q4.chi0.8.er} binaries illustrating the impact of
    eccentricity. In both cases, a genuine BH ringdown cannot be found,
    but the binary with a higher eccentricity exhibits a faster decay of
    the post-merger signal.}
  \label{fig:Q4.chi0.8_ringdown_eccred_vs_no_eccred}
\end{figure}

\section{Impact of eccentricity on the ringdown signal}
\label{sec:ecc_ringdown}

As argued in the main text, reducing the eccentricity represents a
(considerable) additional computational cost that is not strictly
necessary in our study, which is mostly interested in defining the main
features of the dynamics of BHNS mergers and to relate it to QE
considerations. For this reasons, our initial datasets are obtained with
small eccentricities, and subsequently evolved without the eccentricity-reduction
method applied (see discussion in Sec.~\ref{sec:numerical_setup}). The only
exception has been made for the \texttt{Q4.chi0.8} binary, which has
served as a reference to assess the impact of eccentricity on the
dynamics and GW emission. In particular, in Sec.~\ref{sec:mdisk} we have
shown that reducing the orbital eccentricity by a factor of $\simeq 3$
has no significant impact on the properties of the remnant BH and disk,
at least for the eccentricity level we have started with.
On the other hand, in Sec.~\ref{subsec:gw_frequency} we have remarked
that eccentricity results in a significant imprint on the properties of
the waveform during the inspiral and early post-merger. Here, on the
other hand, we illustrate the impact eccentricity has on the post-merger
signal comparing the two realizations of the \texttt{Q4.chi0.8} binary.

Figure~\ref{fig:Q4.chi0.8_ringdown_eccred_vs_no_eccred} provides at a
glance the comparison between the two evolutions of the same binary by
reporting the post-merger signal aligned at the first peak of the
decaying phase. Clearly, none of the two waveforms matches our definition
of BH ringdown discussed in Sec.~\ref{sec:gw_ringdown}. At the same time,
both post-merger signals show an exponential decay that we highlight
using dashed lines and whose corresponding decay times are reported in
the figure.

What can be deduced from these considerations is that for those binaries
that exhibit a clear BH ringdown as a result of the NS plunge or of a
mild disruption, \eg binaries \texttt{Q4.chi0.0}, \texttt{Q7.chi0.8}, or
\texttt{Q7.chi0.8}, a small eccentricity in the initial data does not
spoil the agreement with the perturbative eigenfrequencies. Similarly,
for those binaries that do not exhibit a clear BH ringdown as a result of
a strong disruption, \eg binary \texttt{Q4.chi0.8}, the presence of a
residual eccentricity only mildly affects the properties of the
post-merger signal. Hence, a small residual eccentricity in the initial
data can be tolerated for these two classes of BHNS mergers
for a number of purposes. At the same time, while we
cannot prove it here with our data, we expect that such an eccentricity
will play a particularly important role in those BHNS mergers where a
genuine ringdown is not present, \ie weak-disruption scenarios, but where the post-merger signal has
many of the features of a ringdown, \eg binaries \texttt{Q4.chi0.6} and
\texttt{Q5.chi0.8}. We will explore this conjecture in future work.

\end{document}